\documentclass[apj]{emulateapj}
\usepackage{apjfonts}

\newcommand*{\Msun}{\ensuremath{\mathrm{M_\odot}}}%
\newcommand*{\logMsun}{\ensuremath{\log M/\Msun}}%
\newcommand*{\Mstar}{\ensuremath{M_\ast}}
\newcommand*{\Mg}{\ensuremath{M_{\rm g}}}
\newcommand*{\Mh}{\ensuremath{M_{\rm h}}}
\newcommand*{\Ms}{\ensuremath{M_{\rm s}}}
\newcommand*{\Ns}{\ensuremath{N_{\rm s}}}
\newcommand*{\phig}{\ensuremath{\phi_{\rm g}}}
\newcommand*{\phih}{\ensuremath{\phi_{\rm h}}}
\newcommand*{\phib}{\ensuremath{\phi_{\rm b}}}
\newcommand*{\phif}{\ensuremath{\phi_{\rm f}}}
\newcommand*{\Mstarf}{\ensuremath{M^*_{\rm f}}}
\newcommand*{\Mstarb}{\ensuremath{M^*_{\rm b}}}
\newcommand*{\ab}{\ensuremath{\alpha_{\rm b}}}
\newcommand*{\af}{\ensuremath{\alpha_{\rm f}}}
\newcommand*{\ML}{\ensuremath{M/L}}%
\newcommand*{\Vmax}{\ensuremath{V_{\mathrm{max}}}}%

\begin{document}


\title{The Bimodal Galaxy Stellar Mass Function in the COSMOS Survey to $z\sim 1$: A Steep Faint End and a New Galaxy Dichotomy}

\shorttitle{The Bimodal Galaxy Stellar Mass Function}
\shortauthors{Drory et al.}


\author{N.~Drory\altaffilmark{1}, K.~Bundy\altaffilmark{2,14},
  A.~Leauthaud\altaffilmark{3,4}, N.~Scoville\altaffilmark{5},
  P.~Capak\altaffilmark{5,6}, O.~Ilbert\altaffilmark{7},
  J.S.~Kartaltepe\altaffilmark{8}, J.P.~Kneib\altaffilmark{7},
  H.J.~McCracken\altaffilmark{9}, M.~Salvato\altaffilmark{5,10},
  D.B.~Sanders\altaffilmark{11}, D.~Thompson\altaffilmark{12},
  C.J.~Willott\altaffilmark{13}}

\altaffiltext{1}{Max-Planck Institut f\"ur extraterrestrische Physik, Giessenbachstra\ss e, 85748 Garching, Germany}
\email{drory@mpe.mpg.de}
\altaffiltext{2}{Astronomy Department, 601 Campbell Hall, University of California at Berkeley, Berkeley, CA 94720}
\altaffiltext{3}{Lawrence Berkeley National Laboratory, 1 Cyclotron Road, Berkeley CA 94720}
\altaffiltext{4}{Berkeley Center for Cosmological Physics, University of California, Berkeley, CA 94720, USA}
\altaffiltext{5}{California Institute of Technology, MC 105-24, 1200 East California Boulevard, Pasadena, CA 91125}
\altaffiltext{6}{Spitzer Science Center, California Institute of Technology, Pasadena, CA 91125}
\altaffiltext{7}{Laboratoire d'Astrophysique de Marseille (UMR 6110), CNRS-Universit\'e de Provence, BP 8, 13376 Marseille Cedex 12, France}
\altaffiltext{8}{NOAO, 950 N Cherry Ave, Tucson, AZ 85719}
\altaffiltext{9}{Institut d'Astrophysique de Paris, UMR7095 CNRS, Universit\'e Pierre et Marie Curie, 98 bis Boulevard Arago, 75014 Paris, France}
\altaffiltext{10}{Max-Planck Institut f\"ur Plasmaphysik, Boltzmannstrasse 2, 85748 Garching, Germany}
\altaffiltext{11}{Institute for Astronomy, 2680 Woodlawn Dr., University of Hawaii, Honolulu, Hawaii, 96822}
\altaffiltext{12}{LBT Observatory, University of Arizona, 933 N Cherry Ave, Tucson, AZ 85721}
\altaffiltext{13}{Herzberg Institute of Astrophysics, National Research Council, 5071 West Saanich Rd, Victoria, BC V9E 2E7, Canada}
\altaffiltext{14}{Hubble Fellow}

\slugcomment{ApJ, in press}


\begin{abstract}
  We present a new analysis of stellar mass functions in the COSMOS
  field to fainter limits than has been previously probed at $z \leq
  1$.  The increase in dynamic range reveals features in the shape of
  the stellar mass function that deviate from a single Schechter
  function.  Neither the total nor the red (passive) or blue
  (star-forming) galaxy stellar mass functions can be well fit with a
  single Schechter function once the mass completeness limit of the
  sample probes below $\sim 3 \times 10^{9} \Msun$. We observe a dip
  or plateau at masses $\sim 10^{10} \Msun$, just below the
  traditional $M^*$, and an upturn towards a steep faint-end slope of
  $\alpha \sim -1.7$ at lower mass at all redshifts $\leq 1$. This
  bimodal nature of the mass function is {\em not} solely a result of
  the blue/red dichotomy.  Indeed, the blue mass function is by itself
  bimodal at $z \sim 1$.  This suggests a new dichotomy in galaxy
  formation that predates the appearance of the red sequence. We
  propose two interpretations for this bimodal distribution.  If the
  gas fraction increases towards lower mass, galaxies with $M_{\rm
    baryon} \sim 10^{10} \Msun$ would shift to lower stellar masses,
  creating the observed dip.  This would indicate a change in star
  formation efficiency, perhaps linked to supernovae feedback becoming
  much more efficient below $\sim$10$^{10} \Msun$. Therefore, we
  investigate whether the dip is present in the baryonic (stars+gas)
  mass function.  Alternatively, the dip could be created by an
  enhancement of the galaxy assembly rate at $\sim$10$^{11} \Msun$, a
  phenomenon that naturally arises if the baryon fraction peaks at
  $M_{\rm halo} \sim 10^{12} \Msun$. In this scenario, galaxies
  occupying the bump around $\Mstar$ would be identified with central
  galaxies and the second fainter component of the mass function
  having a steep faint-end slope with satellite galaxies. The low-mass
  end of the blue and total mass functions exhibit a steeper slope
  than has been detected in previous work that may increasingly
  approach the halo mass function value of -2. While the dip feature
  is apparent in the total mass function at all redshifts, it appears
  to shift from the blue to red population, likely as a result of
  transforming high-mass blue galaxies into red ones.  At the same
  time, we detect a drastic upturn in the number of low-mass red
  galaxies.  Their increase with time seems to reflect a decrease in
  the number of blue systems and so we tentatively associate them with
  satellite dwarf (spheroidal) galaxies that have undergone quenching
  due to environmental processes.
\end{abstract}



\section{Introduction}\label{sec:introduction}

Galaxy formation and evolution is believed to be driven primarily by
two processes: firstly, the successive merging of their parent dark
matter halos causing accretion of material and ultimately mergers
between galaxies; and secondly, the feedback-regulated conversion of
gas into stars within galactic disks with subsequent potential
rearrangement of the disk material by dynamical processes (secular
evolution).  Both processes contribute to the growth in stellar mass
of galaxies with time. The stellar mass function of galaxies and its
evolution with time is therefore fundamental to the understanding of
galaxy formation.

The ability to estimate galaxy stellar masses has advanced in recent
years in large part because of increasing access to near-IR
photometry.  Estimates are typically made by fitting multi-band
photometry with stellar population synthesis libraries (see, e.g.,
\citealp{Brinchmann+2000,Bruzual+2003,Drory+2003,Drory+2004,Maraston+2006,Marchesini+2008,Conroy+2009}),
fitting specific spectral features when spectroscopy is available
\citep{Kauffmann+2003}, or the full spectrum when high-quality spectra
are observed \citep{Reichardt+2001,Panter+2004}. So far, only the
photometric fitting technique has been a viable option for
high-redshift surveys. These measurements provide masses with
accuracies of $\sim 0.1-0.3$~dex, and systematic uncertainties of up
to a factor of two depending on the selection and number of
photometric bands included and the assumptions made on, among others,
the shape of the IMF, the allowed star formation histories, the dust
extinction model, or the underlying stellar population synthesis
method (see, e.g.,
\citealp{Drory+2004,Kannappan+2007,Marchesini+2008,Conroy+2009} for
systematic studies on this matter).

Utilizing such techniques, the build-up of the stellar mass density from
redshift $z \sim 6$ to the present epoch has been the subject of several
studies in the past decade, often relying on deep multi-band imaging
surveys in the UV to near-infrared wavelength range
\citep{Brinchmann+2000,Drory+2001a,Cohen2002,Dickinson+2003,Fontana+2003,Rudnick+2003,Glazebrook+2004,Conselice+2005,Chapman+2005,Drory+2005a,Rudnick+2006,Eyles+2007,Grazian+2007,Stark+2007}.

The stellar mass function in the local universe has been measured from
large imaging and spectroscopic surveys such as 2dF, SDSS, and 2MASS
\citep{Cole+2001,Bell+2003,Perez-Gonzalez+2003,Panter+2004,Baldry+2008}.
At distances up to $z \sim 1.5$, a number of groups have established a
picture of the evolution of the mass function with some detail
\citep{Drory+2003,Fontana+2004,Bundy+2005,Borch+2006,Bundy+2006,Arnouts+2007,Pozzetti+2007,Ilbert+2009},
with generally good agreement between different datasets. To some
lesser detail and accuracy, deep surveys have provided data spanning
$0 < z \lesssim 5$
\citep{Drory+2005a,Conselice+2005,Fontana+2006,Yan+2006,Grazian+2007,Elsner+2008,Perez-Gonzalez+2008,Marchesini+2008},
and even some estimates at $z \sim 7$ \citep{Bouwens+2006}. So far,
these high-redshift studies of the stellar mass function emphasized
the evolution of galaxies of mass $\logMsun \gtrsim 10$. Speaking
very broadly, the stellar mass density decreases by a factor of two to
$z \sim 1$, with the most massive galaxies already being in place at
earlier epochs. The evolution appears to accelerate quickly beyond $z
\sim 1.5$.

In this paper, we concentrate on the low-mass galaxies that have
typically been below the completeness limits of previous work at $z >
0.1$. Generally \citet{Schechter1976} fits to the galaxy stellar mass
function with faint-end slope $\sim -1.1$ to $\sim -1.3$ have been
found adequate to describe the galaxy population (even separated
morphologically, by color, or star formation activity;
\citealp{Pannella+2006,Borch+2006,Arnouts+2007,Pannella+2009,Ilbert+2009}). Recently,
though, a steepening of the slope of the luminosity function below
$M_i \sim -17$ in the local universe has been convincingly detected in
clusters
\citep{Driver+1994a,Trentham+2002,Hilker+2003,Popesso+2005,Popesso+2006},
groups \citep{Trentham+2002,Trentham+2005,Gonzalez+2006}, and in the
field \citep{Blanton+2005a}. For example, \citet{Trentham+2002} find
that the luminosity function in the Virgo cluster, in the NGC~1407
group, and in the Coma~1 group is steep between $M_R$ of -18 and -15
(and flattens again only at $M_R > -15$). Moreover,
\citet{Baldry+2008} find that the local galaxy stellar mass function
steepens as well below $\logMsun \sim 9.5$ (but see also
\citealp{Li+2009}).

The steepening of the mass function can also be interpreted as a
bimodality: the mass function consists of a sum of (at least) two
components. This bimodal behavior has now also been detected at
redshifts $z>0.1$. \citet{Pozzetti+2009} find bimodal mass functions to
$z\sim 0.5$ from the zCOSMOS spectroscopic survey. They interpret the
mass function as being composed of early-type galaxies dominating the
massive part and late-type galaxies dominating the less massive part
and contributing the steep faint-end slope. Each of these components
is well fit by a Schechter function. \citet{Bolzonella+2009} use the
same sample to investigate the bimodality as a function of
environment. They find that at $z\lesssim 0.5$, the shape of the galaxy
stellar mass function in high- and low-density environments become
markedly different, with high density regions showing a stronger
bimodality.

We extend the study of the shape of the galaxy stellar mass function,
particularly at low masses, to $z \sim 1$, with stellar mass limits
$\sim 1.5$~dex lower than can be achieved with spectroscopic
studies. We investigate actively star-forming galaxies and passively
evolving galaxies separately; we study how the change in slope may be
caused by the presence of multiple galaxy populations, that taken
together, lead to a mass function shape that is more complex than a
single power law with an exponential cutoff or even a simple
combination of early- and late-type components. We show that the blue
mass function itself is bimodal and that passive galaxies exhibit a
faint-end upturn, likely caused by dwarf spheroidal galaxies linked to
the faint end of the blue galaxy population.

This paper is organized as follows: in \S~\ref{sec:sample} we
introduce the galaxy sample that we use in this work. In
\S~\ref{sec:stellar-mass} we discuss the stellar population models
used to derive stellar masses and the resulting mass completeness
limits. In \S~\ref{sec:mf} we present the stellar mass function of
active and passive galaxies and we discuss our results in
\S~\ref{sec:discuss}. Finally, we summarize this work in
\S~\ref{sec:summary}.

Throughout this work we assume $\Omega_M = 0.3$, $\Omega_{\Lambda} =
0.7, H_0 = 70\ h_{70}^{-1} \mathrm{km\ s^{-1}\ Mpc^{-1}}$. Magnitudes
are in the AB system. We will denote galaxy stellar masses by the
symbol $M$ -- or $\Mg$ where an explicit distinction form halo masses,
denoted by $\Mh$, is necessary. The symbol $M^*$ is reserved for the
characteristic mass parameter of the Schechter function.


\section{Sample Selection}\label{sec:sample}

\begin{figure*}[t]
  \centering
  \includegraphics[width=0.95\textwidth]{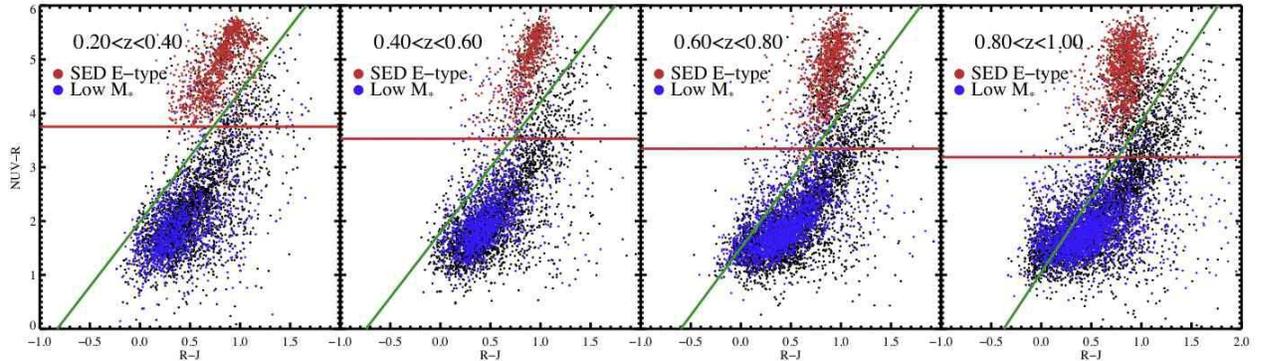}
  \caption{A comparison of the selection techniques of passive
    galaxies in the $NUV\!-\!r$ vs.\ $r\!-\!J$ color-color
    plane. Galaxies identified as passive by SED fitting are marked as
    red points. The lines show the color cuts defined by
    \citet{Williams+2009} for identifying passive systems. In
    addition, galaxies with masses between the mass limit and 3 times
    the mass limit in each redshift bin are marked in blue, to look
    for low-mass galaxies with unusual colors which would indicate a
    possible problem with the photometric
    redshifts.\label{fig:nuv_r_j}}
\end{figure*}

The primary focus of this paper is determining the abundance of
low-mass galaxies and characterizing the shape of the mass function,
especially at the low-mass end.  These galaxies are faint by
definition, with $i \gtrsim 23$ and $K \gtrsim 23$, sources that are
typically beyond the range of magnitude-limited spectroscopic surveys
like DEEP2, VVDS, or zCOSMOS
\citep{Davis+2003,Le-Fevre+2005,Lilly+2007}.  As such, we must rely on
photometric redshifts.

We use the COSMOS catalog with photometric redshifts derived from 30
broad and medium bands described in \citet{Ilbert+2009a} and
\citet{Capak+2007}, version 1.5, dated April 05, 2008. We restrict
ourselves to objects with $i^+_{AB} < 25$. At this limit, the rms
photo-z accuracy at $z<1$ is $\sim 0.03$ in $\Delta z/(1+z)$ (Fig.~9
in \citealp{Ilbert+2009a}). At $z > 1$ the quality of the photometric
redshifts quickly deteriorates.  The detection completeness at
$i^+_{AB} = 25$ is $> 90\%$ \citep{Capak+2007}. Additionally, we
require a detection in the $Ks$ band to ensure that the stellar mass
estimates at the faint end are still reliable, thus we add the
constraint $Ks<24$. The surface brightness sensitivity limit for the
$i^+$ data is 28.4~mag~arcsec$^{-1}$ at 5$\sigma$. At this limit, an
object of 25th magnitude would have to be 5.4 arcsec across (with no
bulge) to have a chance of being missed. At 23rd magnitude that number
increases to 7 arcsec. It is hence very unlikely that surface
brightness is the dominant selection effect at low luminosities where
galaxies are very compact (in contrast to the situation in
local-universe surveys; see simulations in \citealp{Capak+2007}).

To ensure high-quality photometric redshifts we will limit the
redshift range to $z < 1$ and to sample enough volume to $z > 0.2$. We
have made sure that our results are not sensitive to this selection
through repetition of our analysis with relaxed magnitude limits. Our
mass functions do not change shape, nor change their faint-end slope
if we drop the $Ks$-band selection and push the $i^+$ selection to
26th magnitude. Dropping the $Ks$-band selection leads to a gain in
about one bin in depth (0.25~dex in mass), albeit with increased
stellar mass errors by about 20\% on individual sources below the
$Ks$-band limit, but no change in the faint-end slope (the $z$ and
$J$-band still provide enough information on the NIR light). Nor do
our results change if we impose a brighter selection cut of $i^+_{AB}
< 24$, apart from a restriction in redshift range.

We use the photometric data in the $u^*$ (CFHT), $B_J$, $V_J$, $g^+$,
$r^+$, $i^+$, $z^+$ (Subaru), $J$ (UKIRT), and $Ks$ (CFHT) bands for
our analysis, totaling 9 bands. We explicitly exclude the
IRAC1-channel data (3.6~$\mu$m) due to their confusion-limited nature
at $m_{\rm AB}=24$ which would otherwise restrict the analysis to
brighter magnitudes than is possible with the $Ks$ band data.  This is
in contrast to the mass function analysis by \citet{Ilbert+2009}, who
use an IRAC1-selected catalog. However, we have verified that adding
the IRAC1 data where available does not alter the stellar masses
significantly and our mass functions are consistent in the overlapping
mass range with the ones derived by \citet{Ilbert+2009} including the
IRAC1 channel. In a similar analysis \citet{Fontana+2006} find that
including IRAC1-IRAC4 at $z \lesssim 1$ data does reduce the
uncertainties in the mass estimate somewhat, but does not change the
masses systematically (Fig.~1 in their paper).  The reason for this is
that in the redshift range of interest here ($z \lesssim 1$), the
restframe near-IR light is sampled well enough by the $J$ and $Ks$
data, so that the addition of IRAC1 data does not add much information
about the presence of an old stellar population.

\begin{deluxetable}{llclll}[h]
\tablewidth{0pt}
\tablecolumns{6}
\tablecaption{\label{tab:sample}Sample Size}
\tablehead{
\colhead{$z_{\rm min}$} & \colhead{$z_{\rm max}$} & \colhead{Volume} & \multicolumn{3}{c}{Number of Galaxies}\\
\colhead{} & \colhead{} & \colhead{$10^6$~Mpc$^3$} & \colhead{Active} & \colhead{Passive} & \colhead{Total}}\\
\startdata
       0.2 & 0.4 & 0.56 & 27931 & 3553 & 31484\\
       0.4 & 0.6 & 1.23 & 27258 & 2289 & 29547\\
       0.6 & 0.8 & 1.92 & 34298 & 2587 & 36885\\
       0.8 & 1.0 & 2.53 & 37131 & 2954 & 40085\\
\enddata
\end{deluxetable}

In addition to the above magnitude limits, we require that a
photometric redshift be assigned to the objects and that the object
not be in a masked image region (e.g.\ near bright stars, image
border, detector defects) in any of the $B_J$, $V_J$, $i^+$, and $z^+$
bands. The total area of the survey we use after masking is 1.73
square degrees.

We use two methods to remove stars from the sample.  First, sources
where the best-fitting stellar SED template has a lower $\chi^2$ than
the best-fitting galaxy SED template are discarded. This method is
discussed and shown to be very reliable at $i^+ < 24$~mag
(\citealp{Ilbert+2009a}; 2\% of the total population at $i^+ < 24$
could be stars not recognized by the SED classifier while only 0.2\%
of extended sources in ACS are classified as stars by the
$\chi^2$-criterion). In addition, we remove sources with point-like
ACS morphologies using the catalog presented in
\citet{Leauthaud+2007}.  This ACS cut removes an additional 1--3\% of
the total number of sources and is especially important among passive
galaxies at $z > 0.6$. We also note that $\sim 15\%$ of point-like
sources in the COSMOS field are found to be galaxies, however,
\citet{Robin+2007} show that these are faint high-$z$ objects that
will not affect our analysis.

Our final sample contains 138001 sources in the redshift range
$0.2<z<1$. We will divide this redshift range into four equally spaced
bins in the following analysis. Table~\ref{tab:sample} lists the
number of sources in our final sample and its subsamples as well as
the volume probed by each redshift bin.


\subsection{Reliability of the Photometric Redshifts}\label{sec:photoz_reliability}

While our mass functions yield consistently steep faint-end slopes in
all of the redshift bins studied, the lowest redshift bin provides the
deepest probe of the low-mass galaxy population and our strongest
constraints at this mass range.  A top concern, therefore, is that
distant galaxies with much larger redshifts may have been mistakenly
shifted to the $0.2 < z < 0.4$ bin and, with faint apparent
magnitudes, assigned masses with $\logMsun < 9$.  Indeed, photo-$z$
estimates often suffer from degeneracies between the 4000\AA\ break at
$z \sim 0.2$ and the Lyman break at $z \sim 4$, especially if $U$-band
or near-IR photometry is shallow or missing.  We largely mitigate
these problems by requiring detections for all sources in $i^+$ and
$Ks$ in addition to implementing our magnitude limits to reject noise
and false detections.  We confirm this by verifying that our results
do not change if we reject all sources which show a second peak in
their redshift probability distribution with amplitude larger than
5\%.  We note that \citet{Ilbert+2009a} find no evidence of redshift
degeneracies or persistent catastrophic outliers in the COSMOS
photo-$z$ catalog at our magnitude limits that could threaten the
robustness of our result.

Beyond the photometry requirements, we have searched the NUV-R-J
color-color diagram for sources with unusual colors that could
indicate a contaminating population.  This is illustrated in
Fig.~\ref{fig:nuv_r_j}.  Given the mass completeness limit estimate
for each redshift bin, $M_{{\rm lim},i}$, we highlight in blue those
galaxies with $M < 3\times M_{\rm lim}$.  These represent the lowest masses
at which the star-forming sample is still complete.  In large part,
they follow the locus of star-forming galaxies.  No outlier population
is apparent.  We have also repeated this experiment for galaxies whose
photo-$z$ uncertainties are particularly large (a 68\% confidence
interval larger than 0.5 in redshift) and again find no evidence of
contamination. Finally, we use Monte-Carlo simulations described in
more detail below to convince ourselves that the mass function is
sufficiently robust against the errors in the photometric redshifts.


\subsection{Separating Star-forming from Passive Populations}\label{sec:separating}

\begin{figure*}[t]
  \centering
  \includegraphics[width=0.95\textwidth]{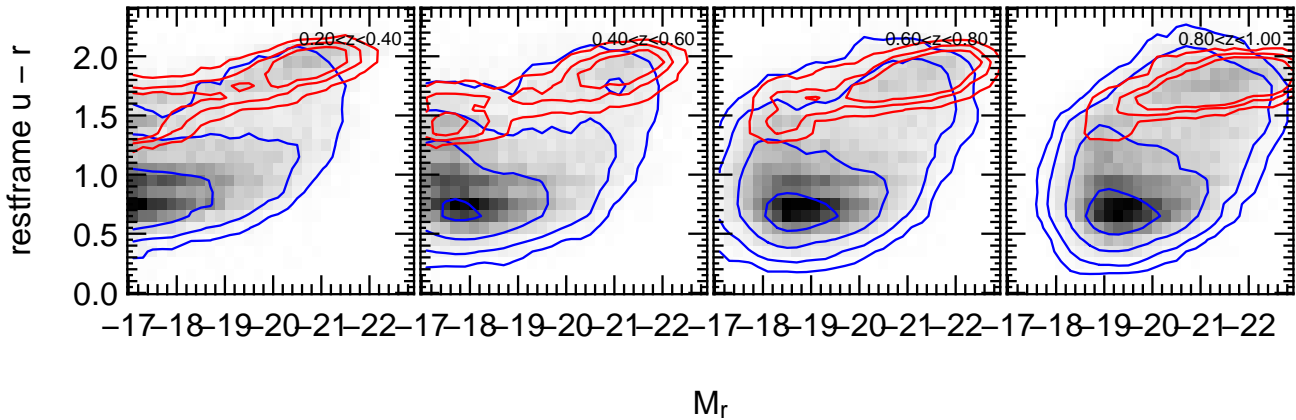}
  \caption{The distribution of galaxies in our sample in the absolute
    magnitude, $M_r$, versus $u\!-\!r$ restframe color plane. The
    greyscale shading marks the density of all objects in this plane,
    while the blue and red contours outline the distributions of the
    blue (active) and red (passive) subsamples,
    respectively. \label{fig:mr-ur} }
\end{figure*}

The low-mass galaxies contributing to the steep faint-end slope of the
mass function are too faint to be morphologically classified using the
available HST data.  Instead, we turn to their colors and SEDs to shed
light on their properties.  Motivated by the ability to break
degeneracies between age and dust, we use a UV-optical-IR (NUV-R-J)
color-color diagram to study our sample, finding that the best-fitting
SED templates returned by the photo-$z$ code do a good job of
distinguishing star-forming from passive galaxies.  We group together
SED types 1--8 from the COSMOS photo-$z$ catalog as ``SED
early-types'' (see \citealp{Ilbert+2009a} and \citealp{Polletta+2007}
for details on the SED templates) and highlight them in red in
Fig.~\ref{fig:nuv_r_j}.  At most redshifts, the early-type SED
designation overlaps well with the associated clump of passive
galaxies identified with red NUV-R colors and blue R-J colors (see
\citealp{Williams+2009} and \citealp{Ilbert+2009a}).  A joint
color-cut indicated by the two lines would select additional passive
systems that are not selected by the SED classification.  At $z
\lesssim 0.6$ the color cuts select $\sim 10\%$ more passive galaxies,
many of these systems scattered between star-forming and passive
sequences.  At $z \sim 1$, however, the SED method identifies as many
as 15\% fewer galaxies in the passive clump.

With none of these classifiers being demonstrably preferable, we
choose to proceed with the SED-based classification. We have
re-analyzed the data with the NUV-R-J-based classification, and find
that the results are fully consistent. In particular, this excludes
that the mass function of faint red galaxies is distorted
significantly by some faint blue galaxies (which are more numerous by
a factor $>10$) being misidentified as red galaxies by one of these
classification methods but not the other. Also, if misclassified blue
galaxies were (partly) responsible for the faint red population, we
would expect this population to become more numerous at high $z$ where
$S/N$ is lower. This turns out not to be the case (see below).

In Fig.~\ref{fig:mr-ur} we show the distribution of the galaxies in
our sample in the absolute magnitude, $M_r$, versus $u\!-\!r$
restframe color plane. The greyscale shading marks the density of all
objects in this plane, while the blue and red contours outline the
distributions of the blue and red subsamples, respectively.


\section{Deriving Stellar Masses}\label{sec:stellar-mass}

Stellar masses (the mass locked up in stars) are routinely derived by
comparing multi-band photometry to a grid of stellar population models
of varying star formation histories (SFHs), ages, and dust content. We
follow the same approach in this work.  Our procedure is described in
detail and compared against spectroscopic and dynamical mass estimates
in Drory, Bender, \& Hopp \citeyearpar{Drory+2004}, although in the
present work we will explore a wider set of stellar population models
described in the following paragraphs.


\subsection{Stellar Population Models}\label{sec:sp-models}

We allow for SFHs consisting of two components: a main population of
stars formed in a smooth SFH and a modulation by a second population
formed in a burst of star formation just prior to observation. The
main component is parameterized by a SFH of the form $\psi(t) \propto
\exp(-t/\tau)$, with $\tau \in [0.5, \infty]$~Gyr and a metallicity of
$-0.6 < \mathrm{[Fe/H]} < 0.3$. The age, $t$, (defined as the time
since the onset of star formation) is allowed to vary between 0.5~Gyr
and the age of the universe at the object's redshift. We linearly
combine this with a starburst modeled as a 100~Myr-old constant star
formation rate episode of solar metallicity. We restrict the burst
fraction, $\beta$, to the range $0 < \beta < 0.15$ in mass. Higher
values of $\beta$ are degenerate and unnecessary since such SEDs are
covered by models with a young main component.  We adopt a Chabrier
\citeyearpar{Chabrier2003} IMF truncated at 0.1 and 100~\Msun\ for
both components. This choice of IMF is described by a power-law at
$M\geq1$~\Msun\ and a log-normal distribution below. The Chabrier IMF
yields masses lower by about 0.25~dex compared to a Salpeter
\citeyearpar{Salpeter1955} single-slope IMF of the form $dn/dm \propto
m^{-2.35}$. We use the stellar population synthesis codes of
\citet[][BC03]{Bruzual+2003} and the BC03 models updated in
2007\footnote[1]{Often referenced as \citet{Charlot+2007}} with an
improved treatment of TP-AGB stars, hereafter referred to as BC07
models.

Additionally, both components are allowed to exhibit a different and
variable amount of extinction by dust. We assume a
\citet{Calzetti+2000} form for the extinction law. This takes into
account the fact that young stars are found in dusty environments and
that the starlight from the galaxy as a whole may be reddened by a
(geometry dependent) different amount. In fact, \citet{Stasinska+2004}
find that the Balmer decrement in the SDSS sample is independent of
inclination, which, on the other hand, is driving global extinction
(\citealp[see, e.g.,][]{Tully+1998}). We verified that using the Milky
Way extinction law or the SMC extinction law instead does not change
our conclusions. This is due to our restricted redshift range which
means we do not probe too far into the restframe UV where the different
slopes of the extinction laws matter most and because low-mass
galaxies at $z<1$ are not heavily obscured, unlike galaxies of the
same mass at $z \gtrsim 2$.

We compute the full likelihood distribution on a grid in this
6-dimensional parameter space ($\tau, \mathrm{[Fe/H]}, t, A_V^1,
\beta, A_V^2$), the likelihood of each model being $\propto
\exp(-\chi^2/2)$. To compute the likelihood distribution of \ML, we
weight the \ML\ of each model by its likelihood and marginalize over
the stellar population parameters. The uncertainty in \ML\ is obtained
from the width of this distribution, which falls between $\pm 0.1$ and $\pm
0.3$~dex at 68\% confidence level on average. The uncertainty
has a weak dependence on mass (increasing with lower $S/N$ photometry),
and much of the variation is in spectral type: early-type galaxies
have more tightly constrained masses than late types.

We have also compared the masses obtained with this code to the ones
obtained with the code employed by \citet{Bundy+2006} and the
\citet{Bruzual+2003} stellar population synthesis models. We find that
the masses computed by both codes are consistent and show no trends
with luminosity or redshift. Hence the stellar masses are not
sensitive to the particular choice of grid parameters and fitting
procedure (within reasonable limits, of course).

Of much graver concern than errors stemming from uncertainties in the
photometry are systematic problems with the stellar population
libraries and the implicit priors on evolutionary histories introduced
by the finite model grid. For a comprehensive analysis see
\citet{Marchesini+2008}. One of the main issues is the treatment of
the thermally pulsing horizontal giant branch (TP-AGB) stars, whose
light dominates in intermediate-age ($0.5-2$~Gyr) populations from the
$I$-band through the near-IR \citep{Maraston2005,Maraston+2006}. The
contribution of these stars can lead to changes of up to a factor of
two to three in the stellar mass estimates for extreme cases (however,
these are unlikely to be observed at $0<z<1$) and if the relevant
wavelength range is poorly covered by the data. TP-AGB spectral
signatures are strongest at $\sim 1~\mu$m rest-frame, which is always
within the observed-frame colors in our dataset. Nevertheless, we base
our model grid on the BC03 models as well as on the BC07 models which
incorporate an improved prescription for the treatment of the TP-AGB
phase (see also \citealp{Bruzual2007}).

Despite the problems with late stellar evolutionary phases,
\citet{Conroy+2009} in a thorough investigation of population
synthesis modeling, argue that the \ML\ ratios are largely resistant
to the uncertain contribution from TP-AGB stars as well as other
limitations in the models. We will nevertheless compare stellar masses
based on both evolutionary synthesis codes to make sure our results do
not depend on a specific choice of a stellar population model.


\subsection{Mass Completeness Limits}\label{sec:mass-completeness}

Determining the selection function in stellar mass for a flux-limited
sample is not possible in a rigorous way: knowledge of the intrinsic
frequency distribution of mass-to-light ratios of the galaxy
population {\em above and below} the flux limit is necessary, but
generally not available. Several methods to (conservatively) {\em
  estimate} the completeness limit have been employed. Most commonly
and simplistically, one assumes that a maximally old population (and
hence with maximally high \ML\ in the absence of dust) at the flux
limit of the survey will yield a conservative upper limit to the mass
that potentially could be affected by incompleteness
(\citealp[e.g.][]{Drory+2003,Drory+2005a,Fontana+2006,Borch+2006,Bundy+2006,Perez-Gonzalez+2008}).

\begin{deluxetable}{llcc}
\tablewidth{0pt}
\tablecolumns{4}
\tablecaption{\label{tab:limits}Mass completeness limits.}
\tablehead{
  \colhead{} & \colhead{} &
  \colhead{Active} & \colhead{Passive}\\
  \colhead{$z_{\rm min}$} & \colhead{$z_{\rm max}$} &
  \colhead{$\log M_{\rm lim}/\Msun$} & \colhead{$\log M_{\rm lim}/\Msun$}}
\startdata
       0.2 & 0.4 & 8.3 & 8.9\\
       0.4 & 0.6 & 8.9 & 9.2\\
       0.6 & 0.8 & 9.2 & 9.8\\
       0.8 & 1.0 & 9.4 & 10.1\\
\enddata
\end{deluxetable}

\citet{Marchesini+2008} employed a different approach: they use
successively deeper datasets to empirically determine their mass
completeness function at shallower flux limits. This approach, while
clearly better than the maximum-\ML\ method is not feasible for
large, uniformly deep surveys such as COSMOS (there are no
significantly deeper datasets observed in the same
passbands). However, it is possible in a limited way using smaller
pencil-beam surveys and we will do so as a sanity check.

We wish to separate our sample into blue, star-forming galaxies and
red, quiescent ones. It is therefore also necessary to determine
separate completeness limits for both populations. We will refrain
from attempting to correct for incompleteness.  Instead, we
conservatively estimate the point from which incompleteness may
significantly affect our sample and limit ourselves to masses larger
than this number.  For the red population, a maximal \ML\ approach
seems sufficient. At the redshifts of interest ($0<z<1$), this
population consists of ellipticals and quiescent spirals with little
evolution in number density with masses $\logMsun \gtrsim 10$
\citep{Bell+2004,Bundy+2005,Pozzetti+2007,Perez-Gonzalez+2008,Ilbert+2009,Williams+2009}. Stronger
evolution and dust-extincted starbursts appear in large numbers at
$z\gtrsim 1.5$
\citep{Fontana+2006,Daddi+2007,Perez-Gonzalez+2008,Ilbert+2009}. In
the local universe most low-mass passive galaxies are dwarf
spheroidals which are very unlikely to contain significant amounts of
dust.  We obtain an estimate for the completeness limit of our sample
at $i^+ < 25$ and $Ks<24$ of $\logMsun = 9$ at $z = 0.3$ and $\logMsun
= 10.1$ at $z=0.9$.

Blue star forming galaxies have much more varied \ML\
values. However, the largest variation is in $\sim L^*$ spirals, with
a large contribution to this variation coming from variable dust
extinction; blue objects fainter than $M_r \sim -20$ at $z \sim 0$
have much more uniform properties \citep{Kauffmann+2003} and in
generally much lower amounts of dust ($A_{H\alpha} < 1$~mag at $\logMsun
 < 9$ compared to $A_{H\alpha} \sim 2 \pm 1.5$ at $\logMsun>10$;
\citealp{Brinchmann+2004}). In accordance with the conclusions reached
by \citet{Kauffmann+2003} in the local universe, the mass-to-light
ratios at magnitudes fainter than $M_r \sim -20$ become roughly
independent of magnitude, converging to $M/L_r \sim 1$. At brighter
magnitudes than $M_r \sim -20$, $M/L_r$ rapidly increases with
luminosity.

We can therefore assume that models with $M/L_r \sim 1$ resemble the
faint galaxy population and a limiting mass computed using a range of
models with values of $M/L_r$ close to or above 1 will yield a
sufficiently robust limit.  Using plausible numbers for a rather
extreme model of a low-mass galaxy with a (light-weighted) age of
$1-2$~Gyr, star formation timescale ($\tau = 5-10$~Gyr), and the
maximum dust extinction found in the local universe for such objects
($A_{H\alpha} = 1$~mag), therefore pushing \ML\ to the high side of
the distribution, we obtain completeness limits roughly 0.4 to 0.7~dex
lower in mass than for the red galaxies. We can verify these limits
using the much smaller but deeper FORS Deep Field (FDF) data
\citep{Drory+2005a}. Comparing the COSMOS mass function at $0.4<z<0.6$
with the FDF mass function bin $0.25<z<0.75$ from \citet{Drory+2005a},
we see that the point where the COSMOS data start to fall below the
FDF data coincides with the mass limit for the blue population as
estimated here (see Fig.~\ref{fig:mf_data} below). The completeness
limits are summarized in Table~\ref{tab:limits}.


\section{The Galaxy Stellar Mass Function}\label{sec:mf}

We compute the galaxy stellar mass function using the \Vmax\ method to
account for the fact that fainter galaxies may not be detectable
throughout a whole redshift bin. Hence, each galaxy in a given
redshift bin contributes to the number density an amount inversely
proportional to the fractional volume of the bin in which the galaxy
would be visible. Given the conservative mass completeness limits
applied in our analysis, we find that the \Vmax-corrections to the
mass function data points above the mass limit are negligible (below
5\% in all bins). Therefore, the data used in the analysis of the mass
function are essentially free of incompleteness corrections.


\subsection{Uncertainties due to Photometric Redshifts}\label{sec:photoz_uncertainties}

Aside from the potential for catastrophic failures (discussed in \S\
\ref{sec:photoz_reliability}), the use of photometric redshifts
implies imprecise distances and therefore imprecise absolute
magnitudes or stellar masses. On the exponential part of the mass
function, uncertainties in the distance will scatter objects
preferentially from lower masses (where objects are exponentially more
abundant) to higher masses. It may also introduce a systematic effect
on the faint-end slope.

\begin{figure}[th]
  \centering
  \includegraphics[width=8cm]{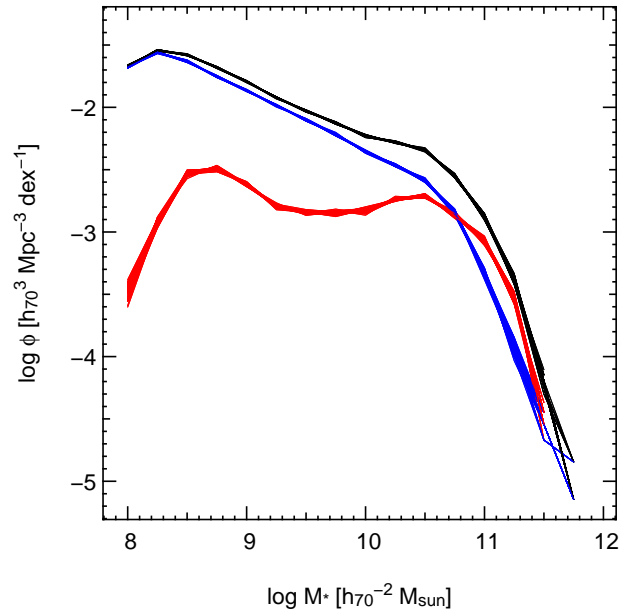}
  \caption{Monte-Carlo realizations of the stellar mass function to
    explore the influence of the uncertainties in the photometric
    redshifts on the shape of the stellar mass function. The mass
    function of passive galaxies is marked in red, that of star-forming
    galaxies in blue, and that of all galaxies in black. Each line
    corresponds to one realization where the redshifts have been
    randomly drawn for each object individually from the 95\%
    confidence region of the redshift probability
    distribution. \label{fig:mf_mc} }
\end{figure}

\begin{figure}[h]
  \centering
  \includegraphics[width=8cm]{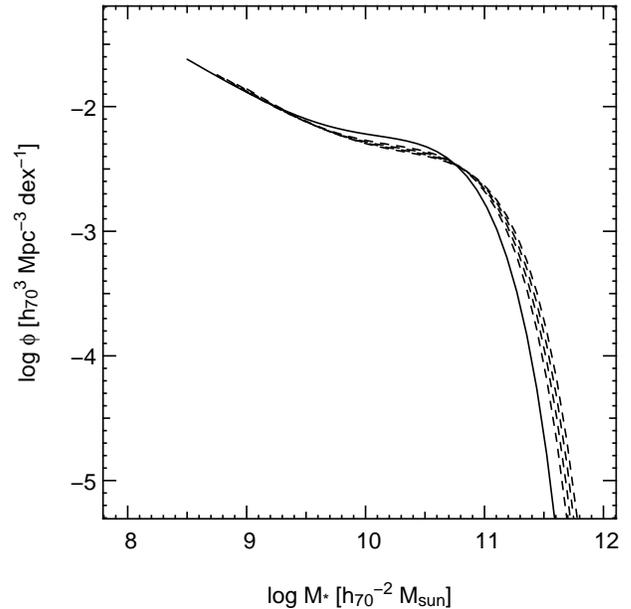}
  \caption{The $z \sim 0$ galaxy stellar mass function from SDSS data
    (\citealp{Baldry+2008}; solid line) convolved with the mass error
    distribution due to the use of photometric redshift based
    distances (dashed lines) at $z \in
    \{0.3,0.5,0.7,0.9\}$.\label{fig:mc_conv}}
\end{figure}

We investigate the effect of uncertainties in the photometric
redshifts on the mass function by means of Monte-Carlo simulations. We
re-assign redshifts to each object drawing from the redshift
probability distribution in the 95\% confidence region around the main
peak. We take the asymmetry of the redshift probability distribution
into account by using different confidence limits for the region below
and above the main peak. We then recompute stellar masses with such
new redshifts and redetermine the mass function. In
Fig.~\ref{fig:mf_mc}, we plot 50 realizations of mass functions
obtained from these simulations. We conclude that redshift errors do
not add large uncertainties at the faint end. The uncertainties in
each mass bin at $\logMsun \leq 10.5$ we derive from these simulations
are $\lesssim 0.03$~dex for the blue and total mass functions, and
$\lesssim 0.05$~dex for the red mass function.  The inferred faint-end
slopes agree to within a fraction of the formal fit uncertainty on the
slope parameter.

The slope may still change systematically, though. To test this, we
run another set of simulations, this time using only objects with
small redshift errors (the best 20\% of the $\Delta z/(1+z)$
distribution) and increase their errors by a factor of 2 and 3. In
neither case is the slope significantly altered compared to the one
obtained with the original small redshift errors (the objects in these
simulations are obviously not quite as faint as those in the whole
sample, but faint enough below $M^*$ to securely measure a change in
slope).

However, as expected, there is an effect on the bright end of the mass
function. Fig.~\ref{fig:mc_conv} shows the $z \sim 0$ SDSS mass
function by \citet{Baldry+2008} and the result of convolving it with
our mass error distributions.  The presence of mass uncertainties
leads to a significant overestimate of the mass function at the
high-mass end.  In summary, we conclude that redshift uncertainties do
not influence the faint-end slope of the mass function systematically
but do add a small random uncertainty.  At the bright end they lead to
larger uncertainties and a systematic ``brightening'' of the
exponential cutoff. We will add the uncertainties from these
simulations (the scatter seen in Fig.~\ref{fig:mf_mc}) to the total
uncertainties in our mass functions.


\subsection{Characterizing the Shape of the Mass Function}

\begin{figure*}[ht]
  \centering
  \includegraphics[width=0.8\textwidth]{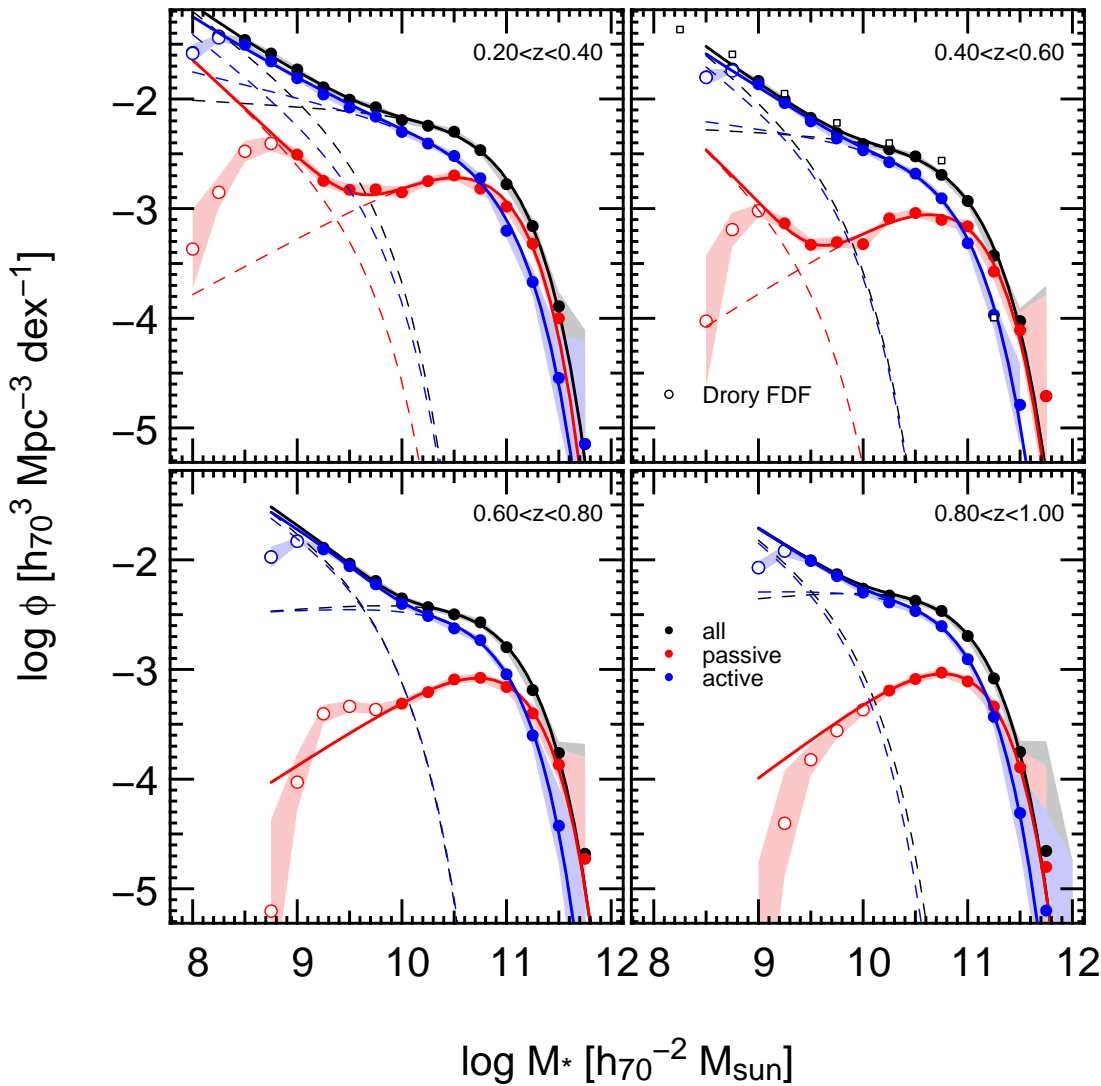}
  \caption{The stellar mass function of galaxies in four redshift bins
    in the interval $0.2 < z < 1.0$. The mass function of passive
    galaxies is marked in red, that of star-forming galaxies in blue,
    and that of all galaxies in black. Data points below the mass
    completeness limits are denoted by colored open symbols. The
    uncertainty in the mass function due to Poissonian errors in the
    counts as well as the uncertainty propagated through photometric
    redshifts (see text) is shown as shaded regions. Data from the
    FORS Deep Field \citep{Drory+2005a} are shown as small black open
    squares at $z=0.5$ for comparison. The solid lines show double
    Schechter function fits to the data (Eq.~\ref{eq:schechter2}). The
    thin dashed lines show the individual bright and faint components
    of the double Schechter functions.\label{fig:mf_data} }
\end{figure*}

We plot the galaxy stellar mass function in Fig.~\ref{fig:mf_data}. The
mass function of passive galaxies is shown in red, that of active
galaxies in blue, and that of all galaxies in black. Filled symbols
mark data points above the formal completeness limit, open symbols data below the
completeness limit and hence where number densities are likely to
be artificially low.

It is evident from Fig.~\ref{fig:mf_data} that a single Schechter
function would not be a good fit to the total mass function (black
points) or to the active galaxy mass function (blue points) at any
redshift $z \leq 1$. At $z \leq 0.5$, the passive galaxy mass function
clearly shows a marked paucity of objects at intermediate mass and a
sharp upturn at low mass. At $z \sim 0.7$ the upturn in the passive
galaxy mass function is still detected, however the mass function is
no longer complete at any point that deviates upwards, and at $z\sim
0.9$ we are unable to detect an upturn. At these two redshift bins,
using only data above the completeness limit, the passive-galaxy data
are adequately fit by a single Schechter function.

One way to describe the mass function shape is as a sum of
(at least) two components, that is, a bimodal distribution.  
Double Schechter functions (the sum of two Schechter functions,
sometimes sharing a common normalization, $\phi^*$, characteristic
scale, $M^*$, or both) have been used by other authors to fit the
luminosity or mass functions with steepening faint-end parts
\citep{Driver+1994a,Popesso+2006,Baldry+2008,Pozzetti+2009}. These work
well to describe the total population and the active
sub-population in our data set. However, more flexibility is needed to
model the transition region between bright and faint populations in the
passive sub-population  at $z<0.6$, as these functions fail to
reproduce the fairly wide dip in the data. We therefore opt to fit all
mass functions by a sum of two Schechter functions without restricting
the parameters further:
\begin{eqnarray}
  \label{eq:schechter2}
  \phi(M)dM & = & \phib(M) dM + \phif(M) dM =\nonumber\\
  & = & \phi^*_{\rm b} \, \left(\frac{M}{M^*_{\rm b}}\right)^{\alpha_{\rm b}} \,
  \exp\left(-\frac{M}{M^*_{\rm b}}\right) dM +\nonumber \\
  & + & \phi^*_{\rm f} \, \left(\frac{M}{M^*_{\rm f}}\right)^{\alpha_{\rm f}} \,
  \exp\left(-\frac{M}{M^*_{\rm f}}\right) dM.
\end{eqnarray}
We define $M^*_{\rm b}$ and $M^*_{\rm f}$ such that $M^*_{\rm b} > M^*_{\rm f}$ thereby
identifying the first term, $\phib(M)$, with a population of bright galaxies and
the second term, $\phif(M)$, with a population of faint galaxies.

The results of fitting Eq.~\ref{eq:schechter2} to the data are shown
in Fig.~\ref{fig:mf_data}. Solid lines mark the fit, $\phi(M)$, to the
active, passive, and total mass functions while the dashed lines show
the individual components, $\phib(M)$ and $\phif(M)$, of each fit. We
again use blue for the active galaxies, red for the passive galaxies,
and black for the total population. The best-fit parameters of
Eq.~\ref{eq:schechter2} are listed in Table~\ref{tab:mffit}. The table
also lists the reduced $\chi^2$ values for the best fit, as well as
the $\chi^2$ obtained by fitting only a single Schechter function to
the data instead of Eq.~\ref{eq:schechter2}. The reduced $\chi^2$
values are between 1 and 2. The best $\chi^2$ values obtained with a
single Schechter function are between 7 and 13, confirming that a
single Schechter function indeed does not provide satisfactory fits
with high significance.

\begin{deluxetable*}{llcccccccc}
\tabletypesize{\scriptsize}
\tablecolumns{10}
\tablecaption{\label{tab:mffit}Double-Schechter fit parameters to the mass function}
\tablehead{%
\colhead{$z$} & \colhead{} & \colhead{$\phi^*_{\rm b}$} & \colhead{$\log \Mstarb$} & \colhead{$\ab$} &
                \colhead{$\phi^*_{\rm f}$} & \colhead{$\log \Mstarf$} & \colhead{$\af$} & \colhead{$\chi^2$} & 
                \colhead{$\chi_{\rm single}^2$\tablenotemark{a}}\\
\colhead{}    & \colhead{} &
                \colhead{$10^{-3}\,h_{70}^3\,\mathrm{Mpc}^{-3}\,\mathrm{dex}^{-1}$} & \colhead{$h_{70}^{-2}\,\mathrm{M_\sun}$} & \colhead{} &
                \colhead{$10^{-3}\,h_{70}^3\,\mathrm{Mpc}^{-3}\,\mathrm{dex}^{-1}$} & \colhead{$h_{70}^{-2}\,\mathrm{M_\sun}$} & 
                \colhead{} & \colhead{} & \colhead{}}%
\startdata
$0.2\dots0.4$&All     &$2.89\pm0.23$&$10.90\pm0.11$&$-1.06\pm0.03$&$1.80\pm0.29$&$9.63\pm0.09$&$-1.73\pm0.09$&1.9&13.0\\
             &Passive &$1.96\pm0.12$&$10.80\pm0.99$&$-0.49\pm0.14$&$0.49\pm0.12$&$9.54\pm0.09$&$-1.85\pm0.20$&1.8&16.8\\
             &Active  &$1.66\pm0.10$&$10.83\pm0.10$&$-1.23\pm0.03$&$0.98\pm0.23$&$9.66\pm0.10$&$-1.75\pm0.12$&1.8&7.0 \\[5pt]
$0.4\dots0.6$&All     &$1.74\pm0.09$&$10.91\pm0.11$&$-1.05\pm0.02$&$1.43\pm0.23$&$9.70\pm0.10$&$-1.76\pm0.16$&1.6&13.6\\
             &Passive &$0.95\pm0.09$&$10.85\pm0.11$&$-0.40\pm0.06$&$0.28\pm0.03$&$9.41\pm0.19$&$-1.84\pm0.29$&1.6&12.7\\
             &Active  &$1.38\pm0.06$&$10.76\pm0.11$&$-1.13\pm0.05$&$1.29\pm0.04$&$9.69\pm0.09$&$-1.71\pm0.19$&1.3&9.6 \\[5pt]
$0.6\dots0.8$&All     &$2.16\pm0.13$&$10.95\pm0.10$&$-0.93\pm0.04$&$2.89\pm0.26$&$9.75\pm0.10$&$-1.65\pm0.08$&1.5&11.9\\
             &Passive &$0.90\pm0.05$&$10.94\pm0.09$&$-0.39\pm0.05$&\nodata       &\nodata       &\nodata        &\nodata&0.7 \\
             &Active  &$1.86\pm0.11$&$10.80\pm0.10$&$-0.95\pm0.04$&$2.78\pm0.31$&$9.75\pm0.10$&$-1.61\pm0.11$&1.6&9.1 \\[5pt]
$0.8\dots1.0$&All     &$2.94\pm0.13$&$10.92\pm0.10$&$-0.91\pm0.03$&$2.12\pm0.29$&$9.85\pm0.10$&$-1.65\pm0.24$&1.4&9.1 \\
             &Passive &$1.03\pm0.05$&$10.91\pm0.11$&$-0.29\pm0.04$&\nodata       &\nodata       &\nodata        &\nodata&0.3 \\
             &Active  &$2.51\pm0.10$&$10.81\pm0.11$&$-0.97\pm0.03$&$2.16\pm0.37$&$9.80\pm0.10$&$-1.66\pm0.36$&1.4&8.0 \\
\enddata
\tablenotetext{a}{Reduced $\chi^2$ for a fit with a single Schechter function.}
\end{deluxetable*}

\begin{figure*}[htpb]
  \centering
  \includegraphics[width=0.95\textwidth]{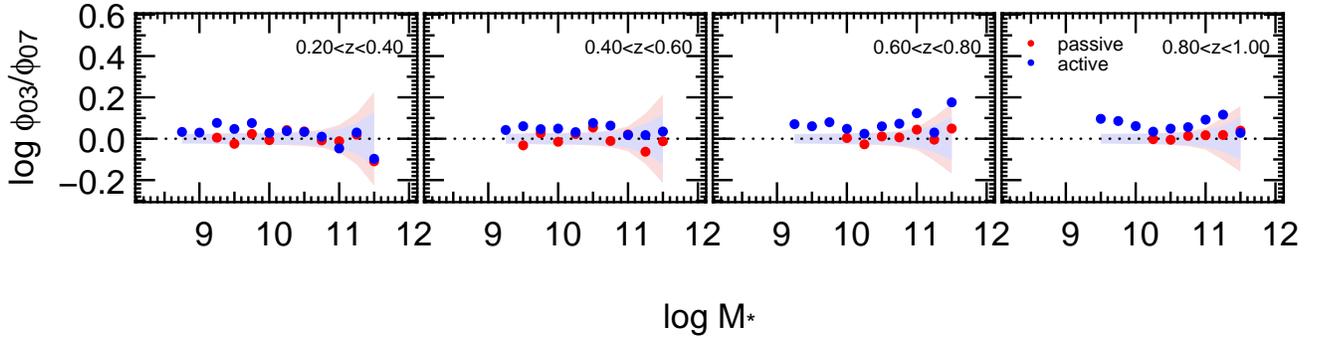}
  \caption{The ratio of the galaxy stellar mass function computed with
    stellar population synthesis models based on the BC03 code and the
    BC07 code which includes a refined treatment of the post-AGB
    stellar evolutionary phases. The values for active and passive
    galaxies are marked by blue and red symbols, respectively. The
    shaded regions denote the statistical uncertainty in the mass
    functions.\label{fig:mf_bc03_bc07}}
\end{figure*}

\begin{figure*}[htbp]
  \centering
  \includegraphics[width=0.95\textwidth]{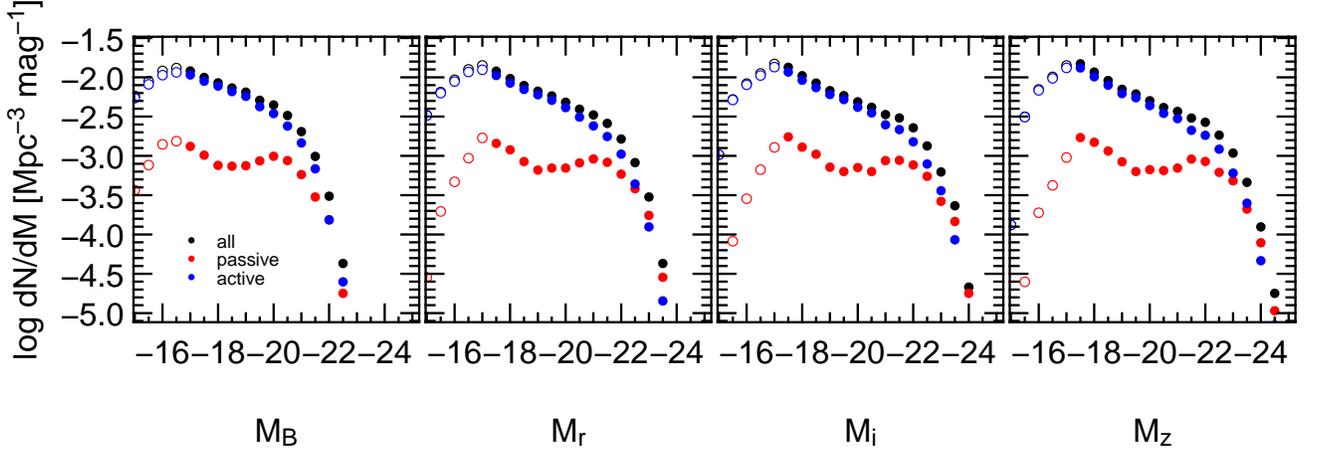}
  \caption{The luminosity function of galaxies (the number density of
    galaxies per comoving volume) in the restframe $B$, $r$, $i$, and
    $z$ bands at $z=0.5$. Passive galaxies are marked in red, star-forming
    galaxies in blue, and all galaxies in black. Data points below the
    completeness limit are denoted by open symbols. \label{fig:lf} }
\end{figure*}

We find a faint-end slope significantly steeper than previous studies,
$\af \sim -1.7 \pm 0.15$ with very little variation with redshift and
between the blue, red, and total samples. With shallower mass limits,
slopes typically around $\alpha \sim -1.2$ are found, (e.g.\
\citealp{Bell+2001,Cole+2001,Drory+2004a,Fontana+2004,Fontana+2006,Ilbert+2009},
although evidence for a steeper slope has been put forward, for
example by \citet{Drory+2008a} from comparing the evolution of the
mass function to the expectation from star formation rates. The value
we find at z=0.3, $\af \sim -1.7 \pm 0.1$, is consistent with that of
\citet{Baldry+2008} in the SDSS, who find a best-fit value of
$\alpha_2 = -1.58 \pm 0.02$, with $\alpha_2 = -1.8$ providing equally
good fits given the non-negligible role of systematic surface
brightness selection effects at their faint end. We find that the
faint-end slope of the blue and the red population at $z=0.3$ and
$z=0.5$, where both can be measured, are remarkably similar. The red
faint-end slope is formally steeper by 0.1, but with a large
uncertainty of $0.2-0.3$. (Evidence for an excess of red dwarf
galaxies was recently also found by \citet{Salimbeni+2008} in the
GOODS data). We do not detect a significant trend of the faint-end
slope of the blue population and hence of the total mass function with
redshift up to $z = 1$.

The characteristic mass of the faint blue sub-component shows a slight
increase with redshift, from $\log \Mstarf \sim 9.6$ at $z=0.3$ to
$\log \Mstarf \sim 9.8$ at $z=0.9$ (significant at the 2-$\sigma$
level), while the characteristic mass in the bright component,
\Mstarb\, does not show evolution in this dataset.

Surveys of $1-2$~deg$^2$ based on single fields can suffer from
significant cosmic variance uncertainties.  The volume probed by our
redshift bins lies between $0.56\times10^6~h_{70}^{-3}$~Mpc$^3$ and
$2.53\times10^6~h_{70}^{-3}$~Mpc$^3$ (equivalent to box sizes between
$82~h_{70}^{-1}$~Mpc and $136~h_{70}^{-1}$~Mpc; see
Table~\ref{tab:sample}). Cosmic variance on such scales is mostly
thought to affect the overall normalization of the mass function,
although possibly the shape as well \citep{Stringer+2009}. In the
COSMOS data, we notice a strong underdensity at $z\sim 0.5$, as has
been seen before (see also \citealp{Meneux+2009}), with the
characteristic density, $\phi^*$, being about a factor of two lower
(more so in the red population than in the blue). Looking at the
distribution of spectroscopic redshifts from zCOSMOS (Fig.~8 in
\citealp{Lilly+2009}), it is evident that over the interval $0.4 < z <
0.65$ there is a marked paucity of dense structures compared to lower
and higher redshifts, which confirms our above observation. The
comparison with the spectroscopic zCOSMOS sample also excludes the
possibility that the density structure in the field is caused by
systematic failures of the photometric redshifts. This variance in
average density influences our ability to make statements on redshift
evolution significantly. Despite the large cosmic variance, we do see
a shift from blue to red galaxies at the bright end as redshift
decreases.

A word of caution is warranted at this point; Eq.~\ref{eq:schechter2}
is by no means the only possible fitting function able to reproduce
the data in a statistically satisfactory way. For instance, a faint
component with a sharper than exponential cut-off works just as
well. Also, the bright part of the (red galaxy) mass function can be
fit by a symmetrically peaked function, similar to a Gaussian. Thus,
one should not base any interpretations solely on the particular fitting
function chosen. We emphasize that the fitting functions in the overlap
region between the bright and faint mass functions should not be taken
to be more than descriptions of the data. Without classifying objects
individually as belonging to the bright or faint subpopulation, the
transition region between the two can be fit in many ways and the
particular representation chosen here should not be assigned physical
meaning.


\subsection{Effects of Stellar Population Models}

Can the features of interest in the stellar mass function be explained
as artifacts introduced by the adopted stellar population models?
First we test whether TP-AGB stars can influence the faint-end slope
of the blue star-forming mass function.  We recompute the mass
function using a stellar population grid based on the BC07 models,
keeping all other model grid parameters identical (see
\S~\ref{sec:sp-models}). Fig.~\ref{fig:mf_bc03_bc07} shows the
difference in the mass function obtained with the BC03 models
(Fig.~\ref{fig:mf_data}) and the mass function obtained with the BC07
models. We do not find significant systematic differences in the mass
function of passive galaxies. In contrast, systematic differences are
small but noticeable in the active population and they increase with
redshift. As expected, the stellar masses obtained with the BC07
models are smaller than the ones obtained with the BC03 models
\citep{Maraston2005,Maraston+2006,Bruzual2007}. The difference is
negligible at $z\sim 0.3$. Beyond $z\sim 0.5$, the difference exceeds
the uncertainties in the masses over much of the mass range,
increasing at masses above and below 10$^{10} \Msun$ to a maximum of
0.1 to 0.15~dex. However, these differences are not large enough to
change the shape of the mass function and the conclusions reached in
this work in any significant way (see also \citealp{Conroy+2009} and
\citealp{Marchesini+2008}).

Could the mass function shape be affected by a failure to correctly
model the very low \ML\ population? We plot the luminosity function of
galaxies in the rest-frame $B$, $r$, $i$, and $z$ bands at $z \sim
0.5$ in Fig.~\ref{fig:lf}. The $B$-band LF shows no bump in the blue
or total galaxy populations; however a bump-like feature appears in
the LF of blue and of all galaxies as one moves to redder filters, and
is unmistakable in the $z$-band LF. So the dip and steep faint-end
slope are present in luminosity functions redwards of the restframe
$B$ band and do not arise from converting luminosity to stellar mass
via SED fitting.


\section{Discussion}\label{sec:discuss}


\subsection{Comparison to Previous Results}

\begin{figure*}[ht]
  \centering
  \includegraphics[width=0.8\textwidth]{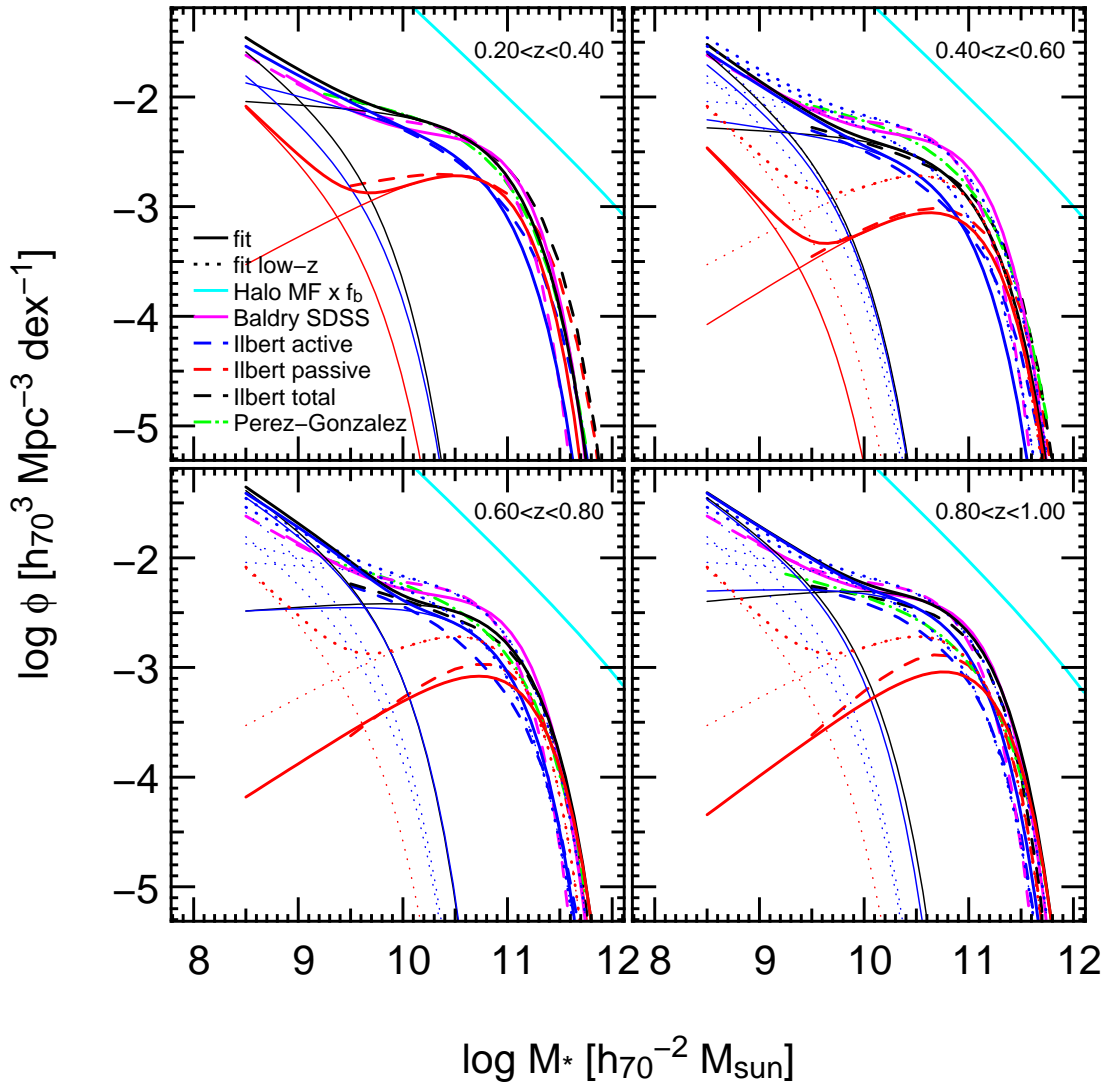}
  \caption{A comparison of the mass function deduced in this work with
    data from the literature. The mass function of passive galaxies is
    marked in red, that of star-forming galaxies in blue, and that of
    all galaxies in black. The solid lines show double Schechter
    function fits to the data (Eq.~\ref{eq:schechter2}; the thin lines
    show the individual bright and faint components of the double
    Schechter function). The fit at low $z$ is repeated in higher-$z$
    panels as dotted lines. The magenta line denotes the
    \citet{Baldry+2008} $z \sim 0$ SDSS-based mass function convolved
    with the photo-z error distributions (see text). The dashed blue,
    red, and black lines are the 3.6$\mu$m-selected mass functions of
    active, passive, and all galaxies, respectively, in the COSMOS
    survey by \citet{Ilbert+2009}. The green dash-dotted line is the
    mass function by \citet{Perez-Gonzalez+2008}. Additionally, we
    plot the Press-Schechter halo mass function scaled by the global
    baryon fraction $f_{\rm b}$ (\citealp{Dunkley+2009}; WMAP5) in
    cyan. \label{fig:mf_lit}}
\end{figure*}

Before discussing possible interpretations of the more complicated
mass function shape apparent in the COSMOS data, it is useful to
compare our results to previous work with the aim of determining
whether evidence for similar behavior has been found in other surveys.
In Fig.~\ref{fig:mf_lit} we compare our mass function to a number of
results from the literature and for later reference we include the
halo mass function \citep{Reed+2007} where the halo masses have been
multiplied by the global baryon fraction, $f_{\rm b} = \Omega_{\rm
  b}/\Omega_{\rm m}$, taken from the WMAP 5-yr data
\citep{Dunkley+2009}. We show the $z \sim 0$ SDSS-based mass function
by \citep{Baldry+2008} convolved with the photo-z error distributions
as shown in Fig.~\ref{fig:mc_conv} and discussed in
\S~\ref{sec:photoz_uncertainties}. We also plot the mass functions in
the redshift range $0<z<1$ of active, passive, and all galaxies in the
COSMOS survey selected at 3.6$\mu$m by \citet{Ilbert+2009} using the
same photometric redshifts as the present work.  Finally the total
mass function selected at $3.6-4.5\mu$m by \citet{Perez-Gonzalez+2008}
is also plotted.

Where they overlap in mass, our mass functions agree very well with
previous work, with some systematic differences stemming from
differences in the underlying stellar population synthesis libraries
and grids of star formation histories (see \citealp{Marchesini+2008}
and \citealp{Conroy+2009} for a systematic study on the influence of
stellar population grid parameters on mass determinations). Of notable
importance, though, is the fact that the bright components in our
two-component fit to the mass function agree very well with the blue
and the red sub-components in \citet{Ilbert+2009}, despite (minor)
differences in the definition of blue and red galaxies. This gives us
confidence that the decomposition of the mass function into a bright
and a faint component with six free parameters is not strongly
degenerate. Specifically, both the mass scale of the bright components
and their faint-end extrapolations are compatible with the single
Schechter function fits to the more restricted data sets. We also
confirm the build-up of the faint part of the red sequence at $z < 1$
observed by other groups
\citep[e.g.][]{Bell+2004,Bundy+2006,Bell+2007,Faber+2007,Perez-Gonzalez+2008,Ilbert+2009,Williams+2009}.

As we discuss further below, the apparent multi-component nature of
the mass function has been discussed at low redshifts by
\citet{Baldry+2008} and \citet{Li+2009} and observed at some level in
other high-$z$ spectroscopic and hence significantly shallower studies
of the COSMOS field \citep{Pozzetti+2009, Bolzonella+2009}.  The
advantage of the current analysis is that it probes to lower mass at
redshifts beyond $z = 0.2$, thus allowing a larger dynamic range which
for the first time at high-$z$ reveals more complicated behavior,
specifically the steepening of the faint-end slope and an additional
population of faint red galaxies (see also \citealp{Salimbeni+2008}).
A similar steepening in the luminosity function below $M_i \sim -17$
has been convincingly detected recently in clusters
\citep{Driver+1994a,Trentham+2002,Hilker+2003,Popesso+2005,Popesso+2006},
groups \citep{Trentham+2002,Trentham+2005,Gonzalez+2006}, and in the
field \citep{Blanton+2005a}. \citet{Baldry+2008} find that the local
galaxy stellar mass function steepens as well below $\logMsun \sim
9.5$ (see also \citealp{Salucci+1999}).


\subsection{Halo Mass to Stellar Mass Relation}

\begin{figure}[hbt]
  \vspace*{0.4cm}
  \centering
  \includegraphics[width=8cm]{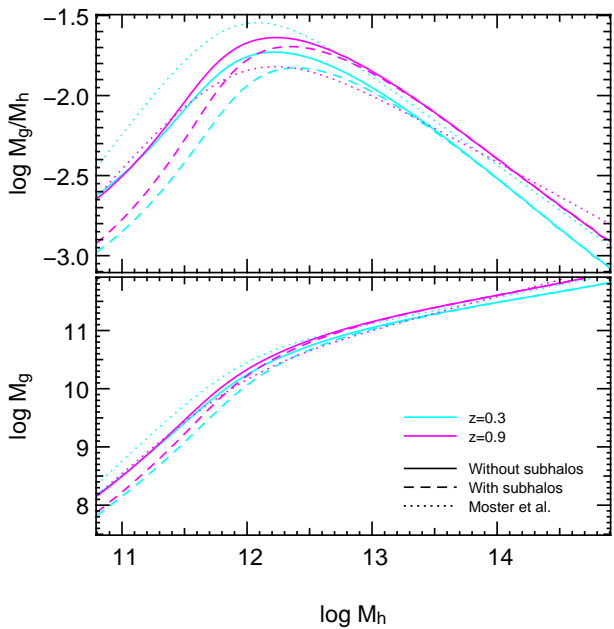}
  \caption{Stellar mass as a function of halo mass (bottom panel) and
    ratio between stellar mass and halo mass (top panel) at redshifts
    of $z=0.3$ and $z=0.9$ determined by abundance matching the galaxy
    stellar mass function to the halo mass function. The relation
    determined by \citet{Moster+2009} are shown for
    comparison. \label{fig:abmatch} }
\end{figure}

For the following discussion of the physical significance of the
various morphological features of the mass function, it is useful to
first place the mass function in the context of the halo mass
function. We will do so by analyzing the halo mass vs.\ galaxy stellar
mass relation, noting that the more complicated shape of the stellar
mass function also leads to a more complicated relation between galaxy
stellar mass, \Mg, and halo mass, \Mh.

The relationship between the halo mass function and the galaxy
(stellar) mass function can be written as
\begin{equation}
  \phig(\Mg,t)=\left|\frac{d\Mh}{d\Mg}\right|\phih(\Mh,t).
  \label{eq:abmatch0}
\end{equation}
Such a one-to-one correspondence between halo mass and galaxy stellar
mass at some time, $\Mh(\Mg)$, can be found by requiring that the
cumulative number density of halos above a given mass and galaxies
above a corresponding stellar mass be equal (abundance matching),
\begin{equation}
  \int_{\Mg}^\infty \phig(M)dM = \int_{\Mh}^\infty \phih(M)dM.
  \label{eq:abmatch}
\end{equation}
This approach has been shown to be sufficiently accurate to match the
two-point correlation function of galaxies and halos
\citep{Conroy+2006,Moster+2009}.

To obtain a more realistic picture of the halo masses of galaxies, we
must include sub-halos in our abundance matching procedure, as halos
(and hence their galaxies) may survive for considerable
time. Analyzing high-resolution cosmological N-body simulations,
\citet{Giocoli+2008} and \citet{Angulo+2008} provide fitting formulae
for the number of sub-halos per host halo per logarithmic interval of
sub-halo to host mass ratio, $d\Ns/d\ln(\Ms/M)$ (see also
\citealp{Gao+2004}). This quantity can be converted to a sub-halo mass
function, $\phi_{\rm sub}(\Ms)$, which is the more convenient quantity
for our purposes by changing variables to obtain $d\Ns/d\Ms$,
multiplying by the abundance of host halos, and integrating over host
halo mass from $\Ms$ to infinity:
\begin{equation}
  \phi_{\mathrm sub}(\Ms) =
  \frac{d\Ns}{d\Ms} =
  \int_{\Ms}^\infty \frac{dN_h}{d\Mh} \frac{1}{\Ms} \frac{d\Ns}{d\ln(\Ms/M)} dM.
  \label{eq:subhalomf}
\end{equation}
This sub-halo mass function can then be added to the distinct halo
mass function to obtain the total mass function that we use to match
to the abundance of galaxies:
\begin{equation}
  \phi(M) = \phi_{\rm distinct} + \phi_{\rm sub}.
  \label{eq:totalhalomf}
\end{equation}

We note that we use the mass of the sub-halo at infall time, not the
mass of the sub-halo at the time of observation for matching
abundances. Most of the galaxy's properties will have been set by the
time infall occurs (and shaped by the potential of the halo at that
time) and galaxy growth is likely to slow down or even stop as soon as
the galaxy becomes a satellite. This approach is motivated by models
of galaxy formation that generically predict that stellar mass is
tightly linked to the potential well in which the galaxy forms
\citep{Kauffmann+1993,Cole+1994,Somerville+1999}. Sub-halos loose mass
by tidal stripping, however their stars are more centrally
concentrated and are not stripped until most of the dark matter has
been lost. The relevant mass scale for sub-halos, therefore, is the
virial mass at infall time.

We proceed using a halo mass function consisting of the distinct halo
mass function by \citet{Reed+2007} and the sub-halo abundance by
\citet{Giocoli+2008} processed through Eq.~\ref{eq:subhalomf}. In
Fig.~\ref{fig:abmatch}, we show the resulting relation between halo
mass, \Mh, and galaxy stellar mass, \Mg at redshifts $z=0.3$ and
$z=0.9$ obtained by abundance matching to our total galaxy stellar
mass function. For comparison, we also show the relation obtained by
\citet{Moster+2009} by populating halos from N-body simulations with
galaxies using semi-analytic methods and the requirement that the
stellar mass function be reproduced.

Without accounting for substructure, $\Mg(\Mh)$ evolves relatively
uniformly at all masses.  With substructure, much less change in the
relation at low mass is seen; there is little evolution in $\Mg(\Mh)$
from $z=0.3$ to $z=0.9$ at $\log \Mh/\Msun \lesssim 11.5$
(corresponding to $\log \Mg/\Msun \lesssim 9.5$, about the mass of the
LMC). This is to be expected if the stellar content of low-mass halos
is limited by feedback. There is evolution in the $\Mg(\Mh)$-relation
at masses above $\Mstar$, where the stellar mass per halo mass
increases by a factor of 1.23 from $z=0.9$ to $z=0.3$. The peak in
star formation efficiency occurs in $\logMsun \sim 12.2$ halos and
reaches 23\% at $z=0.3$ and 19\% at $z=0.9$; galaxy formation -- if
there is a monotonic relation between stellar mass and halo mass such
as the one arrived at by abundance matching -- is always an
inefficient process.

While the flattening of the $\Mg(\Mh)$-relation above $\Mstar$
(corresponding to the exponential cut-off in the galaxy mass function)
is understood as an effect of the inefficiency of cooling in large
halos (e.g.\ \citealp{Rees+1977,White+1978}), the steepening of the
$\Mg(\Mh)$-relation at halo mass of $\logMsun \sim 11$ suggests
another qualitative change in behavior of the feedback efficiency.

The $\Mg(\Mh)$-relation is fundamental to the process of galaxy
formation within dark matter halos and understanding what shapes this
relation is equivalent to understanding the galaxy stellar mass
function. At this point we wish to remark that, due to the derivative
factor $d\Mh/d\Mg$ in Eq.~\ref{eq:abmatch0}, or equivalently, the
integral nature of the abundance matching relation in
Eq.~\ref{eq:abmatch}, a single change of slope of $\Mg(\Mh)$ leads to
a dip-bump structure in the inferred galaxy mass function, given a
power-law halo mass function.  The more abruptly the change of slope
occurs, and the larger it is, the more pronounced the dip-bump feature
in the mass function becomes. The mass function shape we observe,
therefore, leads to an approximately double-power-law form of the
$\Mg(\Mh)$ relation, as can be seen in Fig.~\ref{fig:abmatch}.


\subsection{Interpreting the Shape of the Mass Function}\label{sec:interpretation}

There are several ways to interpret the shape of the total and
type-dependent mass functions in COSMOS.  Broadly speaking, we will
concentrate on two interesting features, namely the ``upturn'' towards
steeper faint-end slopes at low masses ($\logMsun \sim 9$) and an
apparent ``dip'' or plateau at masses just above $\logMsun \sim 10$
that sets in below the traditional $M^*$.  We note that even breaking
down the mass function into these two components is, in itself, an
interpretation since the steepening faint-slope alone may provide a
viable explanation.  As we discuss below, our interest in separately
identifying the dip is motivated by comparisons to the underlying halo
mass function which is a strict power-law at the relevant masses.  In
this context, the dip represents a suppression of the stellar mass
associated with halos in this range; this is a consequence of a change
in $d\Mh/d\Mg$ such that there is a smaller change in the mass of
galaxies per change in the mass of halos. The purpose of this section
is, hence, to explore several physical mechanisms that may explain
both the upturn and the dip and to speculate on the role of future
observations in distinguishing them.

Using the zCOSMOS spectroscopic survey \citet{Pozzetti+2009} interpret
the mass function they recover as being composed of early-type galaxies
dominating the massive part and late-type galaxies dominating the less
massive part and contributing to the steep faint-end slope. They find
that each of these components is well fit by a Schechter function in the
mass range they consider. \citet{Bolzonella+2009} use the same sample to
investigate the bimodality as a function of environment. The find that
at $z\lesssim 0.5$, the shape of the galaxy stellar mass function in
high- and low-density environments become extremely different, with high
density regions showing a stronger bimodality. Also, \citet{Ilbert+2009}
remark that the 3.6$\mu$m-selected COSMOS mass function of all galaxies
is not well-fit by a single Schechter function. They prefer the sum of
their (Schechter) fits to the red and blue galaxies to describe the
whole population, as this reproduces the dip at intermediate mass of
$\log M \sim 10$ and $z \lesssim 0.6$ they observe. Due to the IRAC
selection, they do not reach faint enough limits to detect the steep
faint-end slope or the bimodal structure at higher $z$. We over plot the
data from the FDF \citep{Drory+2005a} at $z=0.5$ in
Fig.~\ref{fig:mf_data} and note that the bimodality is visible there as
well, although \citet{Drory+2005a} do not discuss it because a single
Schechter fit seemed adequate at the time (apart from the data point at
$\log M = 11$ which comes out too high).

We emphasize that we see even more structure in the mass function than
observed in the works discussed above; those studies only detect a dip
in the total mass function and interpret its origin as being due to
the superposition of active and passive populations which by
themselves are unimodal. In this paper, we detect a significant change
in slope in the mass functions of active galaxies as well as a marked
bimodality in the mass function of passive galaxies.

As mentioned above, previous work has attributed the two-component
nature of the mass function to a superposition of a (unimodal) blue
galaxy population modeled as a Schechter function and a red galaxy
population, also modeled as a Schechter function but with a larger
characteristic mass.  In this picture, the bimodality arises through
the transformation of blue galaxies into red galaxies in a process
that must be linked to the mass evolution of a transitioning galaxy to
deplete the dip and create a bump at $M \sim M^*$.  In contrast, we
find that the blue galaxy population by itself shows a dip signature
and a steepening of the power-law slope, which can be interpreted as
arising from a bimodal distribution. In fact, both the red and the
blue populations in Fig.~\ref{fig:mf_data} can be interpreted as
bimodal. However, the bimodality in the blue population seems to be
more pronounced at high $z$ compared to low $z$: it becomes weaker as
redshift decreases. This effect can also be seen in the $\chi^2$
values reported in Table~\ref{tab:mffit}. Indeed, the best $\chi^2$
values for single Schechter fits are obtained for the blue component
at low redshift ($\chi^2_{\rm single} = 7.0$). The signature of the
bimodality (the dip at intermediate mass, or the bump at around $M^*$)
``moves'' from the blue population at high $z$ to the red population
at low $z$.  This is likely because some of the blue galaxies that
make up the bump around $M^*$ in the blue mass function turn red with
time \citep[e.g.][]{Bell+2004,Bundy+2006,Faber+2007,Williams+2009}.

The fact that the blue mass function is itself bimodal at $z \sim 1$
implies that the shape of the total mass function predates the
emergence of the red sequence.  We therefore suggest that the physical
explanation must be a mass-dependent effect or mechanism that is
largely separate from the process that transforms star-forming disks
into passive spheroidals.  In other words, this suggests a new
dichotomy in the galaxy distribution besides the well-known red--blue
distinction.  We will follow this hypothesis in the following, using
the behavior of blue galaxies to stand in for interpretations of the
full population and returning to red systems in the subsequent
section.




\subsubsection{Blue Galaxies}\label{sec:blue}

Before discussing possible physical explanations, we note that among
star-forming, blue galaxies, there is an interesting separation
between (giant) spirals of Hubble types Sa--Scd that dominate the mass
function at the massive end and make up the majority of objects above
the dip, and dwarf galaxies of Hubble types Sd--Sm/Im that dominate
the less massive, power-law part of the blue galaxy mass
function. Disk parameters such as disk radius and luminosity remain
roughly constant (with a large spread) along the Hubble sequence at
types Sa-Scd. For types later than Sd galaxies, the Hubble sequence
turns into a luminosity sequence, with later type morphologies
corresponding to smaller and less massive galaxies (see, for example,
the review by \citealp{Roberts+1994} and references therein).  It is
also worth noting that Disk galaxies without a significant bulge
component exist (e.g., Hubble type Sd and later) and can be accurately
characterized as pure disk systems, harboring at most a nuclear star
cluster.  May we tentatively identify this distinction in disk
properties with the two components in the blue population's mass
function?  If so, the physical processes discussed below must also
account for these morphological differences.

{\bf Star Formation Efficiency.}  The dip in the stellar mass function
corresponds to a change in the stellar mass vs.\ halo mass relation
such that there is a smaller change in the number of galaxies when
measured against the power-law shape of the halo mass abundance
(Figs.~\ref{fig:mf_lit} and \ref{fig:abmatch}).  It is possible that
this deficit arises from a change in the efficiency of star formation
with mass.  In this interpretation, the {\em baryonic} mass function,
if it could be observed, would show a close to power-law shape below
$M^*$, indicating a more smoothly varying fraction of $M_{\rm baryon}
/ M_{\rm halo}$ with $M_{\rm halo}$. For this to still be the case at
later cosmic times we must also require that the fraction of gas
retained by any halo in the interesting mass range must be a smooth
function of halo mass and also that this fraction not be too small.
Then, a decrease in SFR efficiency below a given mass scale (say
$\logMsun \sim 10$) would lead to an increasing gas fraction, $f_{\rm
  gas}$, at lower masses and a {\em decreasing} stellar mass fraction.
This is analogous to the break in the Tully-Fisher relation
\citep{Tully+1977} where galaxies with $V_{\rm c} \lesssim
90$~km~s$^{-1}$ fall below the relation defined by more massive
galaxies. If plotted against total disk mass, $M_{\rm gas}+M_{\rm
  star}$, instead of luminosity, a single relation is restored with
$M_{\rm gas}+M_{\rm star} \sim V_{\rm c}^4$ \citep{McGaugh+2000}.  The
decrease in $f_{\rm gas}$ as a function of $M$ has been
observed. Thus galaxies with baryonic masses near $\logMsun \sim 10$
would have observed $M$ several factors lower, thereby creating a
dip in the stellar mass function at $\logMsun \sim 10$, and leading to
a steepening below $\logMsun \sim 9$.

\begin{figure}
  \centering
  \includegraphics[width=8cm]{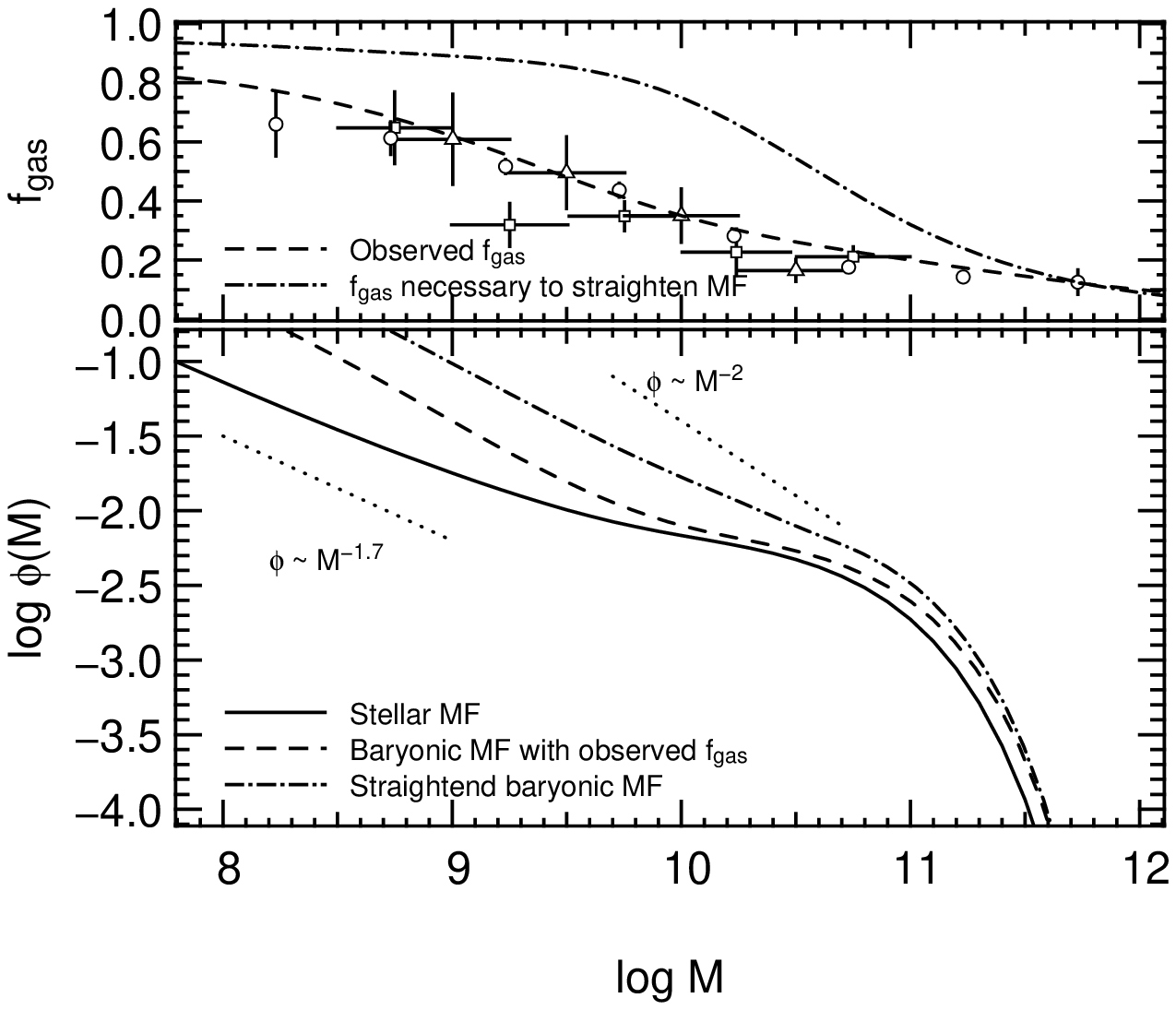}
  \caption{The top panel shows the gas fraction, $f_{\rm gas}$, as a
    function of stellar mass.  The data are taken from a compilation
    by \citet{Hopkins+2009}, including \citet[][
    triangles]{Bell+2001},\citet[][circles]{Kannappan2004}, and
    \citet[][squares]{McGaugh2005}. A spline interpolation to the data
    is shown as a dashed line. The gas fraction necessary to remove
    the dip structure in the mass function is shown as a dash-dotted
    line.  In the bottom panel, we plot the total stellar mass
    function at $z=0.3$ and the baryonic mass function obtained by
    applying the gas fractions from the top panel. Power-law slopes of
    $\phi \sim M^{-2}$ for the halo mass function and $\phi \sim
    M^{-1.7}$ for the stellar mass function are plotted for
    reference. \label{fig:gas_fraction}}
\end{figure}

Using a simple fitting spline to observations of $f_{\rm gas}(M)$
compiled by \citet{Hopkins+2009} including data from
\citet{Bell+2001}, \citet{Kannappan2004}, and \citet{McGaugh2005}, we
transform the total stellar mass function at $z=0.3$ into an
approximate baryonic mass function. The gas fraction hereby is defined
as the ratio of gas mass to stellar plus gas mass, $f_{\rm gas} =
M_{\rm gas}/(M_{\rm gas}+M_{\rm star})$. We plot the results in
Fig.~\ref{fig:gas_fraction}.

Using the observed gas fraction as a function of mass, the dip in the
mass function can be lessened, but not removed. However, as expected,
an appropriate {\em assumption} for $f_{\rm gas}(M)$ leads to
baryonic mass functions in which the dip can be ``straightened'' out
(dot-dash line).  However, the amount of $f_{\rm gas}(M)$ required
is about a factor of 2 higher than current low-$z$ observations
(dashed line).  This comparison involves several large uncertainties,
but plausibly, the difference could be made up for in a warm phase
that is not easily detected \citep{Cen+1999,Dave+2001,Cen+2006}.  One
would also expect $f_{\rm gas}$ to increase with redshift in line with
rising star formation rates.  We are able to conclude that it is at
least {\em possible} that the dip in the stellar mass function
reflects a mass dependence in the ability for baryons to cool and form
stars. 

What causes this mass-dependent SFR efficiency?  We note that the
deviation of the stellar mass function from power-law form at
$\logMsun \sim 9-10$ is close to the scale where supernova feedback
becomes less important, $V_c \sim 100$~km~s$^{-1}$
\citep{Dekel+1986,Benson+2003}. As long as supernova feedback is
efficient in regulating the conversion of gas into stars, the ratio of
stellar mass to halo mass is set by the scaling relation of feedback
efficiency with halo mass (or circular velocity). Once the (DM)
potential is deep enough that supernova feedback can no longer remove
significant amounts of gas from the halo, the attainable ratio of
stellar mass to dark matter mass becomes larger as a larger fraction
of the gas in the halo can cool to form stars. As a consequence, the
stellar mass function deviates upwards from its power-law form. (At
the massive end, the stellar mass is limited by the cooling timescale
becoming too large and the ratio of stellar mass to dark mass
decreases again sharply causing the exponential cut-off in the stellar
mass function \citep{Rees+1977,White+1978}.  The transition from SFR
efficiency suppressed by supernovae feedback to peak SFR efficiency
may therefore lead to the observed dip in the stellar mass function.

There are two potential tests for this scenario.  The most obvious is
to seek better measures of $f_{\rm gas}$ as a function of $M$ and
redshift, but such observations remain challenging.  A second test is
to determine whether the mass scale that defines the dip evolves with
redshift.  Because we expect $f_{\rm gas}$ to be higher in more
massive galaxies at early times (since at $z > 1$ many more are likely
to be rapidly forming stars), the dip in the stellar mass function
should move to higher masses at early times as well.  Unfortunately,
the mass functions presented here are likely to be too affected by
cosmic variance to robustly measure redshift evolution in the dip
feature, but future data sets will be able to overcome this
problem. Also, a more rigorous analysis taking into account
cosmological gas accretion as well as a realistic feedback
prescription and outflow model is needed, but is beyond the scope of
this paper.

{\bf Hierarchical assembly.} Another possibility is that the dip at
$\log M \sim 10$ and the bump around $M^*$ are formed by a depletion of
stellar and gas mass at intermediate mass resulting from a pile-up of
the end-products of mass-dependent assembly around $M^*$.  In this
scenario, both the stellar and baryonic mass functions would exhibit
dips at $\log M \sim 10$ because the assembly of baryonic mass is
accelerated at a certain scale.

It is known that the halo-halo merger rate depends very weakly on mass
\citep{Fakhouri+2008}.  However, if the stellar mass, or in this case,
baryonic mass fraction, peaks in halos at a given mass scale---perhaps
as a result of the SFR feedback discussed above---the resulting merger
rate in terms of stellar mass, $M$, or baryonic mass, $M_{\rm
  baryons}$, can be enhanced near this scale
\citep{Bundy+2009,Stewart2009,Hopkins+2009a} and strengthen the dip
feature.  This is because the dynamical friction timescale (and hence
the merger timescale) depends firstly on the mass ratio of the merging
systems (e.g.\ \citealp{Boylan-Kolchin+2008} and references
therein). Low mass ratio mergers happen only after a long time, and
one-to-one mergers happen quickly. At low mass, one-to-one mergers in
DM correspond to minor mergers in baryons and hence the baryonic
content of a halo assembles more slowly than the halo.  This is due to
the steep relation between stellar mass, \Mg, and halo mass, \Mh,
below $M^*$ (see Fig.~\ref{fig:abmatch}).  At masses near and just
above the scale at which $\Mg(\Mh)$ flattens, more frequent minor halo
mergers can host one-to-one baryonic mergers and thus the baryonic
assembly rates increase markedly.  At even higher mass scales, above
the flattening at $M^*$ of the $\Mh(\Mg)$-relation, the infalling
galaxy is much more likely to become and stay a satellite, even if the
baryonic mass ratio is close to one.  At this point, the baryonic
merger rate drops again.

As a result, low stellar (baryonic) mass galaxies get depleted more
rapidly in one-to-one mergers, and pile-up around the $M^*$ mass scale
($M^* \sim 10^11~\Msun$). This scenario also suggests that the two
populations of galaxies might be thought of as central galaxies
occupying the bump with an increasing fraction of satellite galaxies
at lower masses that possibly form the second, faint, component of the
mass function, leading to a steepening of the low-mass end of the mass
function.  Such satellites in a galaxy or group halo would orbit for a
long time before merging and hence remain visible as individual
galaxies.

If this enhanced assembly scenario were important, it would also
indicate that many major mergers between star-forming (disk) galaxies
must not lead to a final destruction of the disk and truncation of
star formation as the bump signature is already apparent in the blue
galaxy population at $z\sim 1$. Such mergers must produce a remnant
that is still a blue star-forming disk galaxy. Such outcomes of
mergers seem possible if the gas fraction is large
\citep{Springel+2005a,Robertson+2006,Governato+2007,Hopkins+2009b}.
Also, we would expect the dip feature to deepen with time but, in
contrast with SFR scenario above, remain relatively fixed with respect
to mass.

{\bf Steepening Faint-End Slope.} So far we have focused on the dip in
the stellar mass function.  We now turn to the steep slopes at the
low-mass end.  To the extent that this feature is unrelated to the dip
forming process mentioned above, we can consider several possibilities.
To begin with, many authors interpret the faint-end slope as rising to
the maximum set by the halo mass function (with a value of -2).  Such
behavior is expected if feedback processes that regulate the star
formation efficiency become increasingly independent of $M_{\rm halo}$
below some scale.  At low masses, for example, supernova feedback may
so dominate over the ability of halos to retain gas, that the effective
gas or stellar mass fraction becomes a constant, independent of halo
mass.



\subsubsection{Red Galaxies}\label{sec:red}

We now turn to the behavior of the red galaxy mass function.  In the
redshift bins $z=0.5$ and $z=0.3$, where the faint red component of
the mass function is sampled well, the number density of all galaxies
at $M \lesssim 9.5$ does not change much (Fig.~\ref{fig:mf_lit} and
Table~\ref{tab:mffit}). Yet, the number density of faint red galaxies
increases by 0.45~dex, a combined effect of the normalization
increasing from $\phif(z=0.5) = 0.28$ to $\phif(z=0.3) = 0.49$ and the
characteristic mass of the low-mass component increasing from
$\Mstarf(z=0.5) = 9.41$ to $\Mstarf(z=0.3) = 9.54$. This evolution in
faint red galaxies is nearly perfectly mirrored by a decrease in
number of faint blue galaxies. As we have noted above, the faint end
slope of the active and the passive population is consistent with
being equal.

The similarity of the steep faint-end slopes of the active and faint
passive populations and their reciprocal change in number density is
suggestive of the latter originating from the former by shutting off
star formation. It is very tempting to identify these two faint galaxy
populations sharing the same steep faint-end slope with Sm/Im/dIrr
galaxies, and faint spheroidal galaxies, respectively. It is
established that passively evolving dwarf galaxies cluster around
massive galaxies
\citep{Zehavi+2005,Haines+2006,Haines+2007,Carlberg+2009}, and tidal
interactions or ram pressure stripping may well lead to quenching and
some subsequent phase mixing turning Sm/Im galaxies into faint
spheroidal galaxies \citep{Mayer+2001,Grebel+2003,Haines+2007}. Also,
\citet{McCracken+2008} finds that in the CFHTLS in the redshift bin
$0.2 < z < 0.6$ the clustering amplitude for faint red galaxies is
actually higher than that of bright red galaxies, which is to be
expected in our interpretation.



\section{Summary}\label{sec:summary}

Following on previous studies of the stellar mass function in the
COSMOS field \citep{Ilbert+2009, Pozzetti+2009, Bolzonella+2009}, we
present a new analysis of this data set that provides mass functions
to fainter limits than has been previously probed at $z \lesssim 1$.
The resulting increase in dynamic range allows us to characterize and
study features in the shape of the stellar mass function that deviate
from a single Schechter function.  We have tested whether these
features could be introduced by a variety of systematic effects
including both catastrophic photometric redshift errors as well as
increasing photo-$z$ uncertainty at low masses, differences in the way
stellar population models account for TP-AGB stars, and the ability of
stellar mass codes to convert from luminosity to mass.  We conclude
that our results are robust to these effects, although the data are
still limited by cosmic variance.  Our key results follow:

\begin{itemize}

\item Neither the total nor the red (passive) or blue (star-forming)
  galaxy stellar mass functions can be well fit with a single Schechter
  function once the the mass completeness limit of the sample probes
  below roughly $3 \times 10^{9} \Msun$.  We model this more complicated
  behavior using a double Schechter function with 6 free parameters, and
  present the fitting results for 4 redshift bins to $z=1$.  

\item The bimodal nature of the mass function is {\em not} solely a
  result of the blue/red dichotomy.  Indeed, the blue mass function is
  already bimodal at $z \sim 1$.  This suggests a new 2-component model
  for galaxy formation that predates the appearance of the red sequence.

\item We propose two interpretations for this bimodal distribution,
  focusing on the ``dip'' in the blue mass function at $\sim$10$^{10}
  \Msun$.  If the gas fraction increases at lower stellar masses,
  galaxies with $M_{\rm baryon} \sim 10^{10} \Msun$ would shift to
  lower stellar masses, creating the observed dip.  This would
  indicate a change in SFR efficiency, perhaps arising from the
  influence of supernovae feedback which likely sets in below scales
  of $\sim$10$^{10} \Msun$. In this picture, the baryonic mass
  function should not show a dip. Using published (cold) gas fractions
  as a function of stellar mass, we show that cold gas alone is not
  sufficient to eliminate the dip in the baryonic mass function, but
  that the addition of the hard to detect warm gas could potentially
  flatten out the baryonic mass function considerably.

\item Alternatively, the dip could be created by an enhancement of the
  galaxy assembly rate at $\sim$10$^{11} \Msun$, a phenomenon that
  naturally arises if the baryon fraction peaks at $M_{\rm halo} \sim
  10^{12} \Msun$. In this scenario we would identify the galaxies
  occupying the bump around $\Mstar$ with central galaxies and the
  increasing fraction of satellite galaxies at lower mass with the
  second, fainter, component of the mass function and in particular
  the steep faint-end slope.

\item The low-mass end of the blue and total mass functions exhibit a
  steeper slope than has been detected in previous work.  This can be
  interpreted as a {\em steepening} slope, one that may increasingly
  approach the halo mass function value of -2.

\item While the dip feature is apparent in the total mass function at
  all redshifts, it appears to shift from the blue to red population,
  likely as a result of transforming high-mass blue galaxies into red
  ones.  

\item At the same time, we detect a drastic upturn in the number of
  low mass red galaxies.  Their increase with time seems to reflect a
  decrease in the number blue systems and so we tentatively associate
  them with satellite (dwarf spheroidal) galaxies that have undergone
  quenching due to environmental processes.

\end{itemize}

While the broad dynamic range of the COSMOS data set allows us to begin
to characterize some of the more subtle features of the stellar mass
function, the single COSMOS field is still limited by cosmic variance.
Future work over several fields will verify the trends here and shed
light on possible interpretations by constraining how these features
evolve with redshift.


\acknowledgments

We wish to thank R.\ Angulo, R.\ Bender, R.\ Ellis, P.F.\ Hopkins, S.\
Khochfar, and J.\ Tinker for stimulating discussions. We also thank
P.F.\ Hopkins for providing gas mass fractions in electronic form.  We
thank the COSMOS collaboration for granting us access to their
catalogs and we gratefully acknowledge the contributions of the entire
COSMOS team which have made this work possible. More information on
the COSMOS survey is available at {\tt
  http://www.astro.caltech.edu/cosmos}. AL acknowledges support from
the Chamberlain Fellowship at LBNL and from the Berkeley Center for
Cosmological Physics. HJMcC is supported by ANR grant
``ANR-07-BLAN-0228''.



\end{document}